\documentclass[11pt,a4paper]{article}
\pdfoutput=1 
\usepackage[colorlinks=true, linkcolor=black!50!blue, urlcolor=blue, citecolor=blue, anchorcolor=blue]{hyperref}
\usepackage[font=small,labelfont=bf,margin=0mm,labelsep=period,tableposition=bottom]{caption}
\usepackage[a4paper,top=3cm,bottom=2.5cm,left=2.5cm,right=2.5cm,bindingoffset=0mm]{geometry}
\usepackage[T1]{fontenc} 

\usepackage[utf8]{inputenc}
\usepackage[english]{babel}
\usepackage{pifont}
\pdfoutput=1
\usepackage{graphicx,mathtools}
\usepackage{epsfig,cite}
\usepackage{amssymb}
\usepackage{amsmath}
\usepackage{color}
\usepackage{amstext,alltt,setspace}
\usepackage{amsbsy}
\usepackage{array}
\usepackage{booktabs, makecell}
\usepackage{hyperref}
\usepackage{slashed}
\usepackage{url}
\usepackage{geometry}
\usepackage{amssymb,amsthm,cancel,hyperref,graphicx,xcolor}
\usepackage{picinpar,graphicx,xy}
\usepackage[subnum]{cases}
\usepackage{booktabs}
\usepackage{mathrsfs}
\usepackage{amsfonts}
\usepackage{latexsym}
\usepackage{booktabs,graphicx,hyperref,epsfig,multirow}
\usepackage{bm}
\usepackage{comment}
\usepackage{tikz}
\usetikzlibrary{arrows.meta, positioning, fit, calc}
\usepackage{adjustbox}
\usepackage{longtable}

\definecolor{red_cb}{HTML}{e41a1c}
\definecolor{blue_cb}{HTML}{377eb8}
\definecolor{green_cb}{HTML}{4daf4a}
\definecolor{purple_cb}{HTML}{984ea3}
\definecolor{orange_cb}{HTML}{ff7f00}

\definecolor{EmeraldGreen}{HTML}{1ea78d}
\definecolor{EnglishRed}{HTML}{b02427}
\hypersetup{
	pdftitle={Shapley PDFs},
	pdfauthor={Bonnet-Guerrini et al.},
} 

\def\dd{\mathrm{d}}

\newcommand\one{\leavevmode\hbox{\small1\normalsize\kern-.33em1}}
\newcommand{\imag}{\mathrm{i}} 				

\newcommand{\arXiv}[2][]{%
	\ifthenelse{\equal{#1}{}}%
	{\href{http://arxiv.org/abs/#2}{arXiv:#2}}%
	{\href{http://arxiv.org/abs/#2}{arXiv:#2~[#1]}}}

\def\slashchar#1{\setbox0=\hbox{$#1$}           
   \dimen0=\wd0                                 
   \setbox1=\hbox{/} \dimen1=\wd1               
   \ifdim\dimen0>\dimen1                        
      \rlap{\hbox to \dimen0{\hfil/\hfil}}      
      #1                                        
   \else                                        
      \rlap{\hbox to \dimen1{\hfil$#1$\hfil}}   
      /                                         
   \fi}

\newcommand{\tikznode}[2]{%
\ifmmode%
\tikz[remember picture,baseline=(#1.base),inner sep=0pt] \node (#1) {$#2$};%
\else
\tikz[remember picture,baseline=(#1.base),inner sep=0pt] \node (#1) {#2};%
\fi}

\def\mathswitchr#1{\relax\ifmmode{\mathrm{#1}}\else$\mathrm{#1}$\xspace\fi}
\def\mathswitch#1{\relax\ifmmode#1\else$#1$\xspace\fi}

\newcommand{\muF}{\ensuremath{\mu_\mathrm{F}}}

\newcommand{\alphaS}{\ensuremath{\alpha_\mathrm{s}}}

\begin{document}
\begin{flushleft}
\begin{figure}[h]
\includegraphics[width=.2\textwidth]{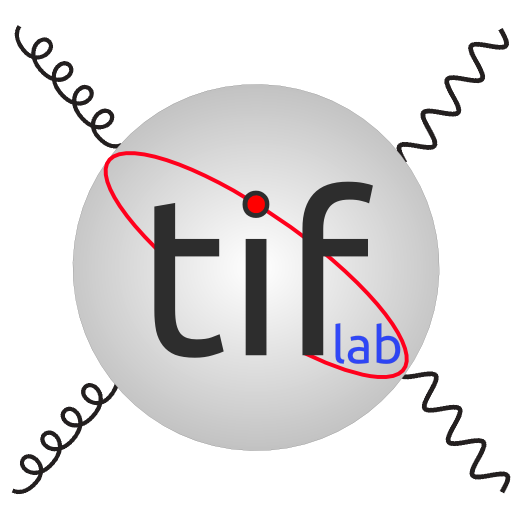}
\end{figure}
\end{flushleft}
\vspace*{-5.0cm}
\begin{flushright}
TIF-UNIMI-2026-5 
\end{flushright}
\vspace*{2.5cm}

\begin{center}{\Large \textbf{{
Interpreting Parton Distributions with Shapley Values
}}}\end{center}
\vspace{1.3cm}
\begin{center}
Rapha{\"e}l Bonnet-Guerrini$^{1,3}$,
Stefano Carrazza$^{2,3}$,  
Stefano Forte$^{2,3}$,\\ 
Eva Groenendijk$^{2,3}$, 
Vincenzo Piuri$^{1}$, 
Ramon Winterhalder$^{2,3}$
\end{center}

\begin{center}
{}$^1$Dipartimento di Informatica, Universit\`a di Milano, 
Via Celoria 18, I-20133 Milano, Italy\\
{}$^2$TIF Lab, Dipartimento di Fisica, Universit\`a di Milano,  
and\\
{}$^3$INFN, Sezione di Milano, Via Celoria 16, I-20133 Milano, Italy

\vspace{0.9cm}


{\bf \large Abstract}
\end{center}

We show that Shapley values can be used to trace how individual parton distributions (PDFs) shape the theory predictions for high-energy observables computed from them.
This provides a tool for assessing the impact of data on PDFs when determining them, and the impact of PDF uncertainties when using the PDFs to compute collider observables.
The Shapley value is computed by treating the regression of PDFs from data as a cooperative game. 
The PDFs are the players, and the reward is the likelihood ($\chi^2$) that characterizes the agreement between data and the predictions obtained from a given PDF, with theory and methodology held fixed. The method is agnostic to the way PDFs have been determined in the first place:  for PDFs  determined with a black-box AI model it may be used in order to explain the behavior of the model, and for PDFs determined using a fixed parametrization it may be used in order to expose the features and potential limitations of the parametrization.  
We find that the method recovers known expectations about which data constrain which PDFs in a global fit, while placing them on a more quantitative footing. 
We demonstrate its effectiveness in two ways. We uncover an unexpected loss of sensitivity of the gluon PDF at intermediate $x$, with potential implications for BSM searches and the gluon fusion Higgs cross section. We also show that the method can be used to improve the hyperparameter optimization procedure currently used by the NNPDF collaboration.

\clearpage

\tableofcontents

\clearpage
\section{Introduction}
\label{sec:intro}

Parton distribution functions (PDFs) enter the computation of every process at a hadron collider, and how precisely they are known directly shapes the reliability of theory predictions.
This is a central challenge for precision particle phenomenology, with two distinct facets.
On the one hand, PDF uncertainties can hamper the search for new physics by masking the signal~\cite{Cole:2026eex}. 
On the other hand, since PDFs are themselves determined from data, understanding how the data constrain them is essential to optimize their extraction.  
This is nontrivial because PDFs are only indirectly related to the data, through perturbative QCD evolution and then convolution with a partonic cross section.

Several tools have been developed to study the sensitivity of PDFs to the data used to determine them~\cite{Wang:2018heo,Jing:2023isu,PDF4LHCWorkingGroup:2022cjn}. 
All of them assume that PDFs are found by maximum-likelihood determination of the parameters of a fixed functional form, and study the dependence of the likelihood on the parameters in the vicinity of its stationary point.
This setup has the disadvantage of only testing PDF sensitivity in the subspace of PDF shapes allowed by the given functional form. Also, it  does not apply when PDFs are determined using a machine learning model such as  neural networks, in which case they are no longer determined by maximum likelihood. The  latter approach to PDF determination was proposed long ago~\cite{Forte:2002fg} and is central to the NNPDF methodology, which produces state-of-the-art sets of parton distributions~\cite{NNPDF:2021njg}.
In this case, the impact of the data on the PDFs is also obscured by the intrinsic opacity of the neural network, whose high-dimensional, non-linear structure makes it difficult to interpret directly~\cite{lipton2016_blackbox}.

Explainable AI (XAI) methods have been suggested in this context as a way to classify PDFs by their sensitivity to individual data points~\cite{Kriesten:2024are}. 
However, many standard XAI techniques that relate features to the behavior of the underlying model~\cite{ribeiro2016whyitrustyou, shrikumar2019learningimportantfeaturespropagating,binder2016layerwiserelevancepropagationneural} are not straightforwardly applicable here: the input is a one-dimensional momentum fraction $x$ rather than a high-dimensional feature vector with clear semantic interpretation, and the quality of the output can only be assessed after propagating the PDFs through the full theory pipeline.

In this work, we address this problem by adapting a specific XAI technique: Shapley values~\cite{shapley_value1953}, a cooperative game-theoretic concept that assigns each player their average marginal contribution across all possible coalitions. 
While Shapley values are often used to elucidate the behavior of a machine-learning (ML) model, such as a neural network, here we use them in a model-agnostic way. 
We take a previously determined set of PDFs as given, without using any information on how it was constructed. We then use the Shapley value to quantify the sensitivity of these PDFs to data, and conversely, the sensitivity of predictions obtained from them to the underlying PDFs. This holds regardless of whether the data and predictions considered here were used in the original PDF determination.
Our method thus allows us to answer questions about the mutual sensitivity of data and PDFs, irrespective of the particular technique used to determine the PDFs.

In ML applications, a common implementation of the Shapley value is the additive SHAP framework~\cite{lundberg2017unifiedapproachinterpretingmodel}.
However, this has been shown to be problematic in high-energy physics applications when correlations are important~\cite{Vent:2025ddm}, which is specifically the case for PDFs.
The reason is that SHAP relies on an additivity assumption, namely that features are approximately independent. We tackle this issue by setting up the problem in such a way that the space of features is low-dimensional, which allows us to compute the Shapley values exactly while accounting for correlations using the sum rules.

Our implementation of the Shapley value views individual PDF flavors as the players of a game. 
The reward of the game is the likelihood ($\chi^2$), computed as follows: starting from a set of PDFs determined with the NNPDF4.0 neural-network model~\cite{NNPDF:2021njg}, we introduce a controlled Gaussian perturbation in a localized range of the momentum fraction $x$, applied to one PDF flavor at a time, for various choices of PDF basis.
Evaluating the Shapley value on different subsets of the global dataset used to determine the PDFs then allows us to disentangle the impact of each subset on each PDF and $x$ range. 

We find that the results for the Shapley value largely reproduce standard lore on the impact of data on PDFs, as discussed in references such as~\cite{Forte:2010dt,Gao:2017yyd}. 
However, we also find that they reveal some less obvious aspects of the connection between PDF features and data, with the possible implications for the possibility  that  PDFs may reabsorb signals of new physics~\cite{Carrazza:2019sec}.
As applications, we discuss a case of diagnostics of lack of sensitivity of PDFs to data that is not revealed by standard methods, and the optimization of the fold design used by NNPDF for $K$-fold cross-validation in their hyperparameter optimization procedure. 

The paper is structured as follows. In Sec.~\ref{sec:PDF-nutshell}, we review the PDF-to-observable pipeline in the NNPDF ML approach, and discuss its black-box nature. 
In Sec.~\ref{sec:shapley-values}, we introduce Shapley values, discuss the limitations of existing frameworks, and describe our exact implementation in PDF space, including the perturbation design and uncertainty quantification. 
In Sec.~\ref{sec:results}, we demonstrate the method on NNPDF4.0, scanning over $x$ regions and individual dataset families, and then discuss examples of applications.
We conclude with a summary and outlook in Sec.~\ref{sec:outlook}.

\section{From PDFs to observables and vice-versa}
\label{sec:PDF-nutshell}

To make our treatment self-contained, we provide a brief overview of the NNPDF approach for PDF determination, focusing specifically on the implementation and methodology used for the construction of the published NNPDF4.0 PDF set. 
As mentioned before, our technique and results do not depend on the particular methodology used to determine the PDFs that we consider. 
However, in addition to helping establish notation and making the paper accessible to readers who are not necessarily familiar with PDF determination, this survey of the NNPDF methodology also partly  motivates the development of our technique. We refer the reader to the NNPDF4.0 paper~\cite{NNPDF:2021njg} and its associated code and documentation~\cite{NNPDF:2021uiq} for a more detailed treatment. 
We first discuss how predictions for physical observables are computed from input PDFs, and then how PDFs are determined from data by solving the corresponding inverse problem in the NNPDF methodology. 
Because we use the published NNPDF4.0 PDF set, we describe settings and assumptions used in that determination, ignoring subsequent improvements and changes in methodology that may have intervened.

\subsection{Factorization and evolution}
\label{sec:factel}

PDFs are determined by performing a regression of theory predictions to experimental measurements of physical observables.
The factorization theorem expresses these predictions as convolutions of PDFs with perturbatively calculable partonic cross sections.
The PDFs characterize the quark and gluon content of the incoming hadrons, while the partonic cross sections describe the scattering of the corresponding initial-state quarks and gluons, collectively denoted as partons.
For processes with two incoming hadrons, this takes the form
\begin{equation}
    \mathcal{O}(Q^2) = \sum_{i,j} \int \dd x_{1}\, \dd x_{2}\;
    f_{i}(x_{1}, \muF^2)\,
    f_{j}(x_{2}, \muF^2)\,
    \hat{\sigma}_{ij}\!\left(x_{1}, x_{2},  \frac{Q^2}{\muF^2}\right),
    \label{eq:factorization_theorem}
\end{equation}
where $\mathcal{O}(Q^2)$ is a physical observable, here the inclusive cross section, depending on a hard scale $Q^2$, $f_i(x,\muF^2)$ is the PDF for flavor $i$ at momentum fraction $x$ and factorization scale $\muF$, and $\hat{\sigma}_{ij}$ is the perturbatively calculable partonic cross section.
The observable does not depend on the factorization scale $\muF$, since this dependence cancels between the PDF and the hard cross section, up to perturbatively subleading terms.
To improve perturbative convergence, the scale $\muF$ is chosen of the order of the physical scale $Q$, so that no large scale ratios arise. After imposing momentum conservation, the integral over $x_i$ takes the form of a convolution of the momentum fraction with some dimensionless ratio of physical scales on which both the observable and the partonic cross section depend. 
For lepton--hadron scattering, only one PDF is present, and the double convolution reduces to a single one. 

The scale dependence of the PDF is fixed by evolution equations of the form
\begin{equation}
\label{eq:evolution}
\frac{\partial f_i(x,\mu^2)}{\partial \ln \mu^2}
=
\sum_{j=1}^{2N_f+1} \int_x^1\frac{\dd z}{z}
P_{ij}\!\left(\alphaS(\mu^2),\frac{x}{z}\right)f_j(z,\mu^2),
\end{equation}
where the sum runs over the gluon and the $N_f$ quark and antiquark flavors with mass $m^2_i\lesssim \mu^2$, and suitable matching conditions are imposed when passing through the production threshold for heavy quarks.
The evolution kernels (splitting functions) in Eq.~(\ref{eq:evolution}) admit a perturbative expansion in powers of the strong coupling
\begin{equation}\label{eq:pertp}
P_{ij}\!\left(\alphaS(\mu^2),x\right)= \frac{\alphaS(\mu^2)}{2\pi} P^{(0)}_{ij}\!\left(x\right)+ 
\left(\frac{\alphaS(\mu^2)}{2\pi}\right)^2 P^{(1)}_{ij}\!\left(x\right)+\dots,
\end{equation}
and give the probability of parton $j$ to emit a parton $i$ that carries a fraction $x$ of its momentum. 
Equation~(\ref{eq:evolution}) implies that Mellin moments
\begin{equation}
\label{eq:mellin}
\bar f_i(N,\mu^2)=\int_0^1 \dd x\;x^{N-1}\,f_i(x,\mu^2)
\end{equation}
of the PDFs satisfy ordinary differential evolution equations driven by anomalous dimensions $\gamma_{ij}(N)$ related by Mellin transformation to the splitting functions: 
\begin{equation}
\label{eq:evolutionn}
\frac{\partial \bar f_i(N,\mu^2)}{\partial \ln \mu^2}
=
\sum_{j=1}^{2N_f+1} \gamma_{ij}\!\left(\alphaS(\mu^2),N\right)\bar f_j(N,\mu^2),
\end{equation}
where
\begin{equation}
\label{eq:mellin_splitting}
\gamma_{ij}\!\left(\alphaS(\mu^2),N\right)=\int_0^1 \dd x\;x^{N-1}\, P_{ij}\!\left(\alphaS(\mu^2),x\right),
\end{equation}

It follows that, even if PDFs must be evaluated at a variety of different scales $\mu$ to obtain accurate theory predictions, they can all be expressed in terms of a set of basis PDFs evaluated at a reference scale $\mu=Q_0$. 
By choosing a suitable (scale-independent) basis of linear combinations of PDFs, the evolution equations decouple into a pair of coupled equations, and $2N_f-1$ decoupled equations. We refer to this basis as the evolution basis. The two coupled equations couple the gluon and the singlet combination, defined as
\begin{equation}\label{eq:singlet}
\Sigma(x,\mu^2)=\sum_{i=1}^{N_f} q^+_i(x,\mu^2), 
\end{equation}
where
\begin{equation}\label{eq:qpm}
q_i^\pm= q_i\pm \bar q_i,
\end{equation}
and $q_i$ and $\bar q_i$ are the quark and antiquark PDFs, respectively. The other decoupled equations evolve specific linear combinations of quark and antiquark distributions, and have at most three independent evolution kernels -- only one at leading order (LO) and two at next-to-leading order (NLO) -- so several equivalent choices of evolution basis are possible.

At all scales, PDFs satisfy a set of integral constraints, namely the momentum sum rule
\begin{equation}
\label{eq:mom_sumrule}
    \int_0^1 \dd x\, x\left(g(x,\mu^2)+ \Sigma(x,\mu^2)\right)=1,
\end{equation}
and the valence sum rules
\begin{equation}
\label{eq:val_sumrule}
    \int_0^1 \dd x\, u^-(x,\mu^2) = 2\int_0^1 \dd x\, d^-(x,\mu^2)=2,\qquad  \int_0^1 \dd x\, s^-(x,\mu^2) = \int_0^1 \dd x\, c^-(x,\mu^2)=0,
\end{equation}
where $u$, $d$, $s$ and $c$ denote the up, down, strange and charm quark PDFs, respectively.

\subsection{PDF parametrization and determination}
\label{sec:pardet}

In the NNPDF approach of Ref.~\cite{NNPDF:2021njg} the PDFs, viewed as a function of $x$ but at a fixed scale $Q_0$, are parametrized by a preprocessed neural network
\begin{equation}
\label{eq:PDF}
  x f_i(x, Q^2_0) = A_i\, x^{1-\alpha_i}(1-x)^{\beta_i}\, \text{NN}_i(x),
\end{equation}
where $\alpha_i$ and $\beta_i$  are preprocessing exponents, and $A_i$ are normalization constants that may be fixed to enforce the sum rules Eqs.~(\ref{eq:mom_sumrule}-\ref{eq:val_sumrule}). PDFs at all scales are obtained by solving the evolution equations~\eqref{eq:evolution}, and then convoluting the PDF at the scale of the data with the appropriate partonic cross sections according to Eq.~\eqref{eq:factorization_theorem}. This provides a map from PDF space to data space, which in the NNPDF methodology is implemented through a suitable  discretization and optimization pipeline~\cite{Barontini:2023vmr}.

The neural network is then trained to minimize a loss function, which takes the form of a negative log-likelihood describing the agreement between theory and data,
\begin{equation}
  L_{\text{NNPDF}} = \chi^2(f,\mathcal{D})= \sum_{n,m}^{N_{\text{dat}}}
  \left[ \mathcal{O}_{n}(f) - d_{n} \right] \Sigma_{nm}^{-1}
  \left[ \mathcal{O}_{m}(f) - d_{m} \right],
  \label{eq:loss_simp_pdf}
\end{equation}
where $\mathcal{O}_{n}(f)$ denotes the theoretical prediction for data point $n$ computed using a given PDF set $f$, $d_{n}$ is the corresponding experimental measurement of the full dataset $\mathcal{D} = \{d_1,\dots,d_{N_{\text{dat}}}\}$, and $\Sigma_{nm}$ is a covariance matrix that encodes both statistical and systematic experimental uncertainties as well as theory uncertainties originating from approximations in the theory predictions.  

The loss minimization is stopped via cross-validation: the data are divided into training and validation samples, the network is trained by minimizing the loss on the training sample, and training is halted once the loss evaluated on the validation sample has failed to improve for a specified number of iterations (the patience). 
PDF uncertainties are determined in two steps. First, a Gaussian distribution of pseudodata replicas $\mathcal{D}^k$
is generated, whose mean and covariance over the replica sample reproduce the original data and their correlated uncertainties. Then, a best-fit PDF $f^k$ is determined for each pseudodata replica, leading to a distribution of best-fit neural networks whose mean and covariance represent the final PDF determination and its associated correlated uncertainties. The training and validation partition is determined randomly for each replica, and the preprocessing exponents are varied randomly in a self-consistently determined interval that ensures that their choice does not bias the final result. 

Both the architecture and other features of the neural network (such as the initialization or the activation function) and the minimization algorithm and its features (such as the maximum gradient or the stopping patience) are determined through a $K$-fold hyperparameter optimization procedure, performed by dividing the data into four folds, excluding one fold at a time from the training process, and optimizing the agreement with the excluded fold.
Note that whereas in the NNPDF approach PDF uncertainties are natively represented  through Monte Carlo replicas, commonly available tools~\cite{Watt:2012tq} enable the conversion of PDF uncertainties represented as a multivariate Gaussian in parameter space into the Monte Carlo replica representation. The Monte Carlo representation is accordingly commonly used to compare and combine different PDF sets~\cite{PDF4LHCWorkingGroup:2022cjn}.

While any set of basis PDFs can be chosen, the basis of the gluon and individual quark and antiquark flavors (flavor basis) as well as the aforementioned evolution basis are of particular interest. In the NNPDF4.0 determination the basis includes eight independently determined PDFs: the charm and anticharm PDFs are assumed to coincide (so  the charm valence sum rule in Eq.~\eqref{eq:val_sumrule} is automatically satisfied), and heavier flavors (bottom and top) are not independently parametrized but rather determined perturbatively. 
The basis PDFs then are, in the flavor basis,
\begin{equation}\label{eq:bfldef}
\mathcal{B}_\text{flav}=\{u,\bar u,d,\bar d, s, \bar s, c=\bar c, g\},
\end{equation}
and in the evolution basis
\begin{equation}\label{eq:bevdef}
\mathcal{B}_\text{ev}=\{V,V_3, V_8,T_3,T_8,T_{15},\Sigma, g\},
\end{equation}
where $\Sigma$ is defined in Eq.~(\ref{eq:singlet}), the total valence distribution is
\begin{equation}\label{eq:valdef}
V=\sum_{i=1}^{N_f} q^-_i,
\end{equation}
and the non-singlet combinations are
\begin{align}
T_3&=u^+-d^+, & T_8&=u^++d^+-2 s^+, & T_{15}&=u^++d^+ +s^+-3c^+,\label{eq:tdef}\\
V_3&=u^--d^-, & V_8&=u^-+d^--2 s^-.\label{eq:vdef}
\end{align}

\subsection{The black-box problem}
\label{sec:black_box}

We can view the set of experimental data (about 4000 for NNPDF4.0) as a probability distribution over a high-dimensional data space, typically a multivariate Gaussian about the central data points, with width determined by the eigenvalues of the total covariance matrix along its eigenvectors. The regression from data to PDFs is an inverse problem that maps the probability distribution in data space onto a probability distribution in PDF space. Because of the way predictions are obtained from PDFs (discussed in Sec.~\ref{sec:factel}), this inverse problem 
has strong non-localities and correlations, and makes the inference chain behave like a black box as described below.
\begin{description}

\item[Discrete data coverage.]  Data provide information at a discrete set of points, moreover in a limited kinematic region. The neural net interpolates and extrapolates this discrete information, so that distinct PDF shapes can yield the same value of the loss function. The space of acceptable shapes, and their likelihood, are determined by the hyperparameters of the neural network and its minimization in such a way that the most likely PDF shapes in the final distribution are generally not those leading to the lowest loss, which usually correspond to  overfitted results. 

\item[Convolutional structure.] Each observable is sensitive to PDFs that are  evaluated not at a single $x$ value, but  rather in the whole region $x_0<x<1$, for some $x_0$ depending on the process kinematics, through the  convolution 
in Eq.~\eqref{eq:factorization_theorem}. The convolution becomes a local ordinary product upon Mellin transformation. Hence the interpolation and extrapolation mentioned above actually happen in a conjugate Mellin space.

\item[Scale dependence.] PDFs are determined at a reference scale $Q_0$, but the  PDFs that enter the physical prediction  Eq.~\eqref{eq:factorization_theorem} are evaluated   at the factorization scale $\mu_F$, in turn chosen of order of a physical scale for each data point. The two are related through convolution with an evolution kernel that solves the evolution Eq.~\eqref{eq:evolution}, thereby further modifying in a scale-dependent way  the relation between the $x$ value of the PDF and the kinematics of the data.
   
\item[Scaling violations.] Because data are available for several values of scale, and the scale dependence (scaling violation) of the PDF must satisfy the evolution Eq.~\eqref{eq:evolution}, the PDFs are determined not only by the data directly, but also by the scale dependence of the data which is related to the PDF by convolution with the splitting function. Hence the PDF is effectively determined by the mutual relation of data at different scales.

\item[Flavor mixing.] Individual quark flavors are a linear combination of different evolution eigenstates, hence they mix with each other as the scale varies, and consequently the linear combinations that appear in a given physical observable are related through this scale-dependent mixing to the flavor basis PDFs. Even if one chooses the evolution basis, observables cannot be fully  expressed in terms of scale-independent evolution eigenstates because the  singlet and gluon satisfy coupled equations with scale-dependent coefficients.

\item[Non-local correlations.] Both data and PDFs that enter different individual predictions of the form of Eq.~\eqref{eq:factorization_theorem} are correlated with each other in a highly non-local way. The PDFs are correlated because of sum rules 
Eqs.~(\ref{eq:mom_sumrule}-\ref{eq:val_sumrule})  that relate PDFs at all $x$ values. The data are correlated through the covariance matrix $ \Sigma_{nm}^{-1}$ that enters the definition of the loss Eq.~(\ref{eq:loss_simp_pdf}), which may correlate  data coming from independent experiments, but sharing a common source of uncertainty.
 
\end{description}

Taken together, these effects make it difficult to answer the simple question, which PDF flavor, and in which $x$ region, is responsible for the theory prediction to be compared to a specific data point. 
These effects are intrinsic to the PDF-to-observable pipeline itself, and are  present regardless of the way the PDFs are determined. 
The use of ML adds a further layer of opacity on top of this. Because the regression is performed by a neural network rather than a fixed functional form, the map between data and PDFs has no analytic form to begin with, compounding the difficulty just described. 
Indeed, an alternative possibility, adopted by several groups~\cite{PDF4LHCWorkingGroup:2022cjn}, is to determine the PDFs by assuming a fixed functional form whose parameters are determined by maximum likelihood. 
In such a case, the mapping between the probability distribution in data space and that in PDF space is known analytically, as it is fixed by the maximum-likelihood condition combined with error propagation, and its behavior in the vicinity of the maximum. 
The price to pay is that only the space of functions allowed by the chosen model is accessible.
In the NNPDF approach the neural network model does not restrict the space of allowed functional forms, but the set of more likely functional forms is effectively chosen by the ML model through the hyperoptimization of its architecture and training parameters and is consequently opaque.

\section{Shapley values for PDFs}
\label{sec:shapley-values}

Motivated by the black-box structure discussed in Sec.~\ref{sec:black_box}, we approach the problem using the Shapley value, a game-theoretic concept~\cite{shapley_value1953} that has more recently been adopted as an XAI technique to probe the behavior of ML models. 
We use it here in its original, broader sense. In conventional XAI applications, the players are the input features of a model, such as a neural network, and the Shapley value explains how that model maps those inputs to its output. Here, instead, the players are the PDFs themselves, already determined by a previous, possibly opaque fitting procedure. We use the Shapley value to relate these PDFs to the theory predictions obtained from them, without ever examining the model that produced the PDFs in the first place. 
While the black-box nature of the NNPDF methodology motivates our approach to understanding the relation between data and PDFs, our solution to the problem is completely independent of the methodology used to determine the PDFs.

\begin{figure}[b!]
\centering
\vspace{1pt}
\input{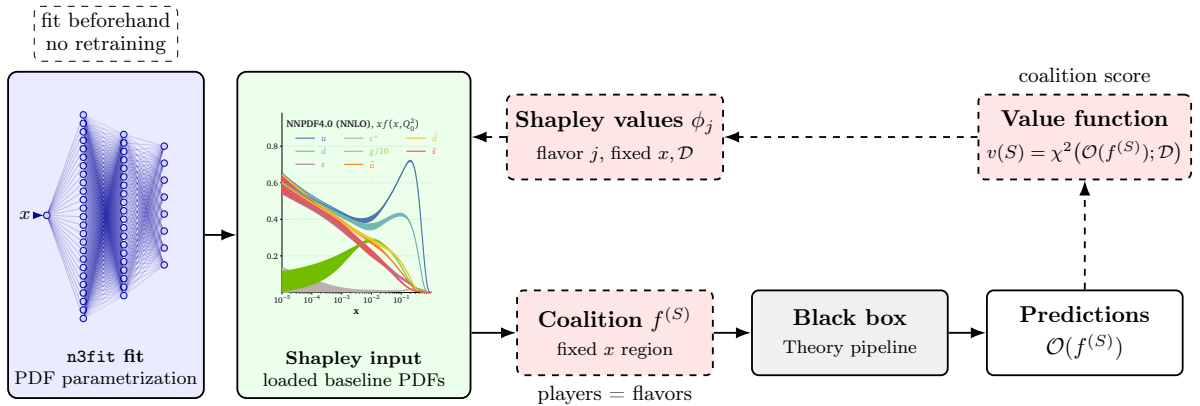}
\caption{Schematic overview of the Shapley value analysis. A previously determined PDF set is loaded as input.
For a fixed $x$ region and coalition $S$, perturbed PDFs $f^{(S)}$ are propagated through the theory pipeline to obtain predictions that are compared to some dataset $\mathcal{D}$ through a suitable $\chi^2$ log-likelihood. Shapley values $\phi_j$ quantify the contribution of each flavor for a chosen $x$ region and dataset $\mathcal{D}$.}
\label{fig:sv4pdf_pipeline}
\end{figure}

Our main idea is to  treat the PDF-to-observable pipeline as a cooperative game, where the players are the individual PDFs and the payoff is the loss. We first briefly review the game-theoretic foundations of Shapley values and its use as an XAI tool, and then we present our specific implementation in the context of PDF determination. Our procedure is schematically summarized in Fig.~\ref{fig:sv4pdf_pipeline}.

\subsection{The Shapley value in game theory and XAI}

In game theory, the Shapley value is the average marginal contribution $\phi_i$ of a player $i$, determined by comparing the outcomes of a cooperative game with and without $i$: $v(S \cup \{i\})$ versus $v(S)$, where $v$ is a value function that maps a coalition, i.e.\ a subset of the total set of $N$ players $S \subseteq N$, to a real number that gives the payoff achieved by that coalition~\cite{shapley_value1953}. 
Note that for brevity we denote with $S$ and $N$ both the sets, and the number of elements in the respective sets. 
The Shapley value is then given by
\begin{equation}
  \phi_i = \sum_{S \subseteq N \setminus \{i\}} w(S) \left[ v(S \cup \{i\}) - v(S) \right],
  \label{eq:comb-shapley-values}
\end{equation}
where the sum runs over all possible coalitions of $N-1$ players not including the $i$-th player, and the weight $w(S)$ ensures that coalitions are weighted accounting for the combinatorial factor that provides the number of different coalitions, namely the reciprocal of the number of different coalitions with $S$ members. 
Indeed, the weight $w(S)$ is given by the number of different (non-ordered) subsets of $S$ out of $N-1$ elements, which is the binomial coefficient times the number $N$ of possible choices of the $i$-th element, 
\begin{equation}\label{eq:weight}
  w(S) = \left[N\left(\begin{array}{c} N-1\\S\end{array}\right)\right]^{-1}=\frac{S! \, (N - S - 1)!}{N!}.
\end{equation}

In ML applications, the Shapley value is used to determine the contribution of individual features to the loss function: the features are the players, and the loss is the value  function. In this context,  the exact computation of Shapley values is usually intractable, since the typical number $N$ of features is of order of 50 or more, hence the sum over $2^{N-1}$ coalitions is unfeasible. A common way out is the use of the
SHAP framework~\cite{lundberg2017unifiedapproachinterpretingmodel} (SHapley Additive exPlanations), that addresses this by unifying a class of additive feature attribution 
methods -- including LIME~\cite{ribeiro2016whyitrustyou}, DeepLIFT~\cite{shrikumar2019learningimportantfeaturespropagating}, and LRP~\cite{binder2016layerwiserelevancepropagationneural} -- and 
approximating the Shapley values by assuming feature independence, which then allows for a linear expansion, such that the number of coalitions only grows linearly, rather than exponentially with $N$. In  high-energy physics applications, where input features are typically strongly correlated, this leads to distorted results, as explicitly demonstrated for jet substructure variables in top tagging in Ref.~\cite{Vent:2025ddm}. As discussed in Sec.~\ref{sec:black_box}, in our case PDFs and their features are also strongly correlated, thereby suggesting that this approximation would be equally problematic. As we shall see, we will consider a setup in which the players are the $N_\text{flav}=8$ PDF flavors  $\mathcal{B}_\text{flav}$ or $\mathcal{B}_\text{ev}$ Eqs.~(\ref{eq:bfldef}-\ref{eq:bevdef}). Hence, even though we will consider a large number of different games, in which one flavor is perturbed in a specific region, for each of them we will only have to consider $2^7=128$ coalitions, thereby enabling the computation of the exact Shapley value Eq.~(\ref{eq:comb-shapley-values}).

\subsection{Shapley implementation for PDFs}
\label{sec:setup}

The idea of our approach is to take as a basic feature the behavior of each PDF in a certain $x$ region.  We start with a pre-existing PDF set, and we consider as a player a perturbed PDF, namely a PDF from this set whose behavior in a given $x$ region has been modified. We then assess the effect of the perturbation by computing the Shapley value for the perturbed PDF in coalition with all other possible coalitions of perturbed PDFs, using as a value function the loss Eq.~(\ref{eq:loss_simp_pdf}) used for PDF determination. 

The players are the $N_\text{flav}=8$ flavors in $\mathcal{B}_\text{flav}$ or $\mathcal{B}_\text{ev}$, see Eqs.~\eqref{eq:bfldef}--\eqref{eq:bevdef}. For any subset $S\subseteq\mathcal{B}_\text{flav/ev}$, a coalition
configuration is obtained by perturbing only the flavors in $S$,
\begin{equation}
\label{eq:coaldef}
f^{(S)}_j =
    \begin{cases}
      f_j^\text{pert}, & j\in S,\\
      f^{(0)}_j,       & j\notin S,
    \end{cases}
\end{equation}
where $f^{(0)}$ denotes the unperturbed baseline PDF set.
The value function is obtained from the loss Eq.~\eqref{eq:loss_simp_pdf}, evaluated using, in each instance in the sum over coalitions, the PDFs corresponding to coalition $S$ as defined in 
Eq.~(\ref{eq:coaldef}). The perturbation is designed in such a way as  to single out the features that we want to study, namely the behavior of each PDF flavor in a given $x$ region. 

\begin{figure}[t!]
    \centering
    \includegraphics[width=\linewidth]{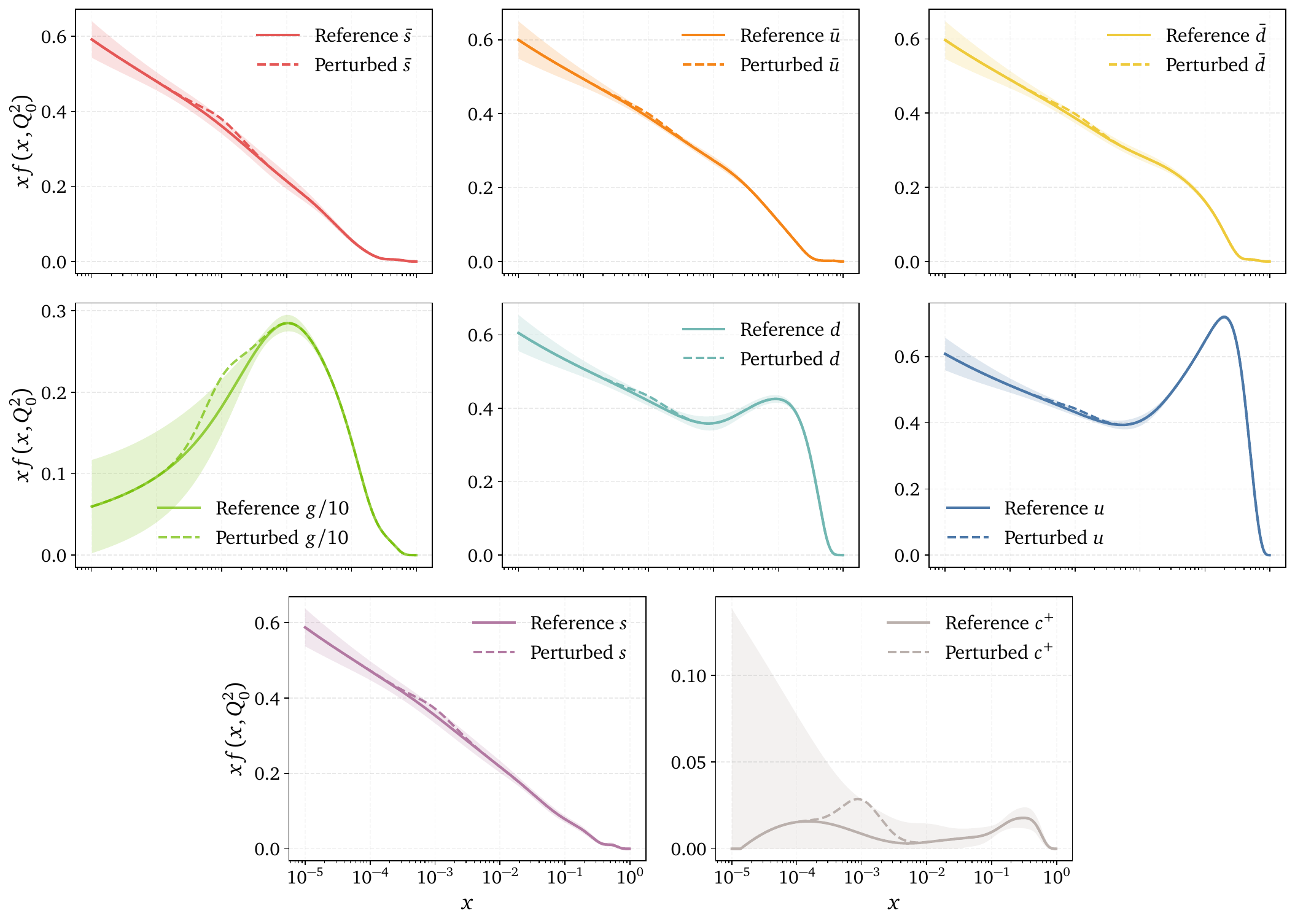}
    \caption{Calibrated perturbation Eq.~(\ref{eq:perturbation-cal}) centered at $x_\mu=10^{-3}$ in
    the flavor basis, with amplitude normalized to $1\sigma$ at $Q_0=1.65~\mathrm{GeV}$ and $w=0.3$.}
    \label{fig:10-4calibrated-perturbation}
\end{figure}

We now discuss the perturbation and corresponding value function in turn. 
In order to select the behavior in a given $x$ region, we apply a localized perturbation, chosen to be a Gaussian bump, thereby ensuring that the smoothness of the unperturbed PDFs as a function of their argument is preserved. Because we want to single out the effect of modifying the PDF about its central value in either direction, the perturbation is applied with either a positive or a negative
sign. 
The bump is centered at some $x=x_\mu$ and has a width $w$ on a logarithmic scale, such that the PDF is probed on a fixed-width interval on a log scale.
The height of the bump is calibrated to the PDF uncertainty at the center of the perturbation, so that the perturbation is physically meaningful and
comparable across flavors. 
Specifically, the height of the positive (negative)
perturbation is set equal to the upper (lower) 68\% confidence interval of the replica ensemble evaluated at $x_\mu$. We thus have
\begin{align}
\label{eq:perturbation-cal}
    xf_{j}^{\text{pert}\pm}(x,Q^2_0) &= xf_j(x,Q^2_0)+\delta^\pm_j(x),
\\\label{eq:deltapm}
    \delta^\pm_j(x) &= \pm\Delta^\pm_j(x_\mu)
    \exp\!\left[-\frac{\bigl(\log_{10}(x)-\log_{10}(x_\mu)\bigr)^2}{2w^2}\right],
\end{align}
where $\Delta^+_j(x_\mu)$ and $\Delta^-_j(x_\mu)$ are the upper and lower 68\% confidence interval bounds relative to the central value of the $N_{\mathrm{rep}}$ replicas.
A representative example of this calibrated perturbation is shown in Fig.~\ref{fig:10-4calibrated-perturbation}.

The value function is then constructed out of the loss in the following way. First, either the perturbation $\delta^+$ or $\delta^-$ Eq.~(\ref{eq:deltapm}) is applied to all PDFs in coalition $S$ Eq.~(\ref{eq:coaldef}). Next, the sum rules Eqs.~(\ref{eq:mom_sumrule}-\ref{eq:val_sumrule}) are enforced in the perturbed PDF set: this means that if the coalition $S$ includes some but not all of the PDFs that enter one of the sum rules, then only the PDFs in the coalition are perturbed according to Eq.~(\ref{eq:perturbation-cal}), but all the other PDFs entering the sum rule are adjusted by modifying their normalization constants $A_i$ in 
Eq.~(\ref{eq:PDF}) in such a way that the sum rule still holds. This then produces for each coalition two perturbed PDFs $f^{(S+)}$ and $f^{(S-)}$, in which the PDFs in the coalitions are perturbed, and the sum rules always hold. Finally, the value function is evaluated by taking the average of the  loss   Eq.~\eqref{eq:loss_simp_pdf} evaluated for these perturbed PDF sets
\begin{align}
\label{eq:value-function}
    v(S)=\frac{1}{2}\left[\chi^2\!\left(f^{(S+)};\mathcal{D}\right)+\chi^2\!\left(f^{(S-)};\mathcal{D}\right)\right],
\end{align}

The Shapley value is finally computed using this value function in its expression Eq.~(\ref{eq:comb-shapley-values}), thereby leading to a value $\phi_j(x_\mu)$, depending on the perturbed flavor $j$ and the center $x_\mu$ of the perturbation.
As mentioned, a common way of representing PDF uncertainties, specifically used by NNPDF, is to deliver PDF sets as sets of replicas $f^k$ such that the central expected PDF is the ensemble mean and the uncertainty and correlation are found from the ensemble covariance over the replica sample. Accordingly, we can determine  both the PDF Shapley value and its uncertainty for a given PDF set by evaluating the Shapley value  $\phi_i^{(k)}= \phi_i(f^k)$ for each PDF replica $f^k$, i.e.\ evaluating  the value function Eq.~(\ref{eq:value-function}) with fixed data $\mathcal{D}$ and  the $k$-th PDF replica $f^k$. The Shapley value $\phi_i$ and its uncertainty $\sigma_{\phi_i}$ are then determined as the mean and  $1\sigma$ standard deviation of $\phi_i$ across the PDF replica ensemble,
\begin{align}
\phi_i\equiv\bar{\phi}_i &= \frac{1}{N_\text{rep}}\sum_{k=1}^{N_\text{rep}}\phi_i^{(k)},
    \qquad
    \sigma^2_{\phi_i} = \frac{1}{N_\text{rep}-1}
    \sum_{k=1}^{N_\text{rep}}\bigl(\phi_i^{(k)}-\bar{\phi}_i\bigr)^2.
    \label{eq:uq}
\end{align}
For a quicker evaluation of the Shapley value, without an evaluation of its uncertainty, we will also compute it directly for the central expected PDF, obtained by averaging over the PDF replica sample.

With our definition, the Shapley value is effectively the marginal change of the value function Eq.~(\ref{eq:value-function}) when flavor $j$ is added to an existing coalition $S$, with a baseline given by  $v(\emptyset)=\chi^2(f^{(0)};\mathcal{D})$, which is just the value of the loss which is found when comparing the unperturbed PDF set $f_j$ under investigation to the data $\mathcal{D}$. 
Accordingly, the Shapley value can be computed for a given PDF set using any dataset: it then measures the sensitivity of the PDF set to the data, which are not necessarily the ones used to determine the PDFs  in the first place. As such, it can be used to assess the impact of adding new data to an existing PDF determination. On the other hand, because it is defined as the impact of a deviation of the given PDF set from its reference best fit, if computed over a subset of the data  used to determine the given PDF set, it assesses how much of the information on the given PDF set came from that data.

Evaluating the value may lead to three possible situations:
\begin{itemize}
  \item $\phi_j(x_\mu)>0$: the perturbation of flavor $j$ increases $\chi^2$ on average, indicating that the dataset $\mathcal{D}$ is sensitive to and constrains that flavor in the given $x$ region.
  
  \item $\phi_j(x_\mu)\approx 0$: the dataset is insensitive to flavor $j$ in the given $x$ region, either because the flavor does not contribute meaningfully to the relevant predictions or because its effect cancels across coalitions.

  \item $\phi_j(x_\mu)<0$: the perturbation improves $\chi^2$ on average, suggesting that the baseline PDF is locally suboptimal for this dataset, which might happen for instance if there are tensions between the data.
\end{itemize}

As a final comment, we note that what allows us to take into account all correlations between features is the small number of coalitions that enables an exact computation of the Shapley index. However, this is only possible because we have defined the coalition of perturbed PDFs according to Eq.~(\ref{eq:coaldef}), in which the same perturbation is applied to all the PDFs in the coalition, so the number of players is equal to $N_{\rm flav}$, and moreover we are only considering symmetric perturbations according to Eq.~(\ref{eq:value-function}). 

These are somewhat restrictive assumptions, because, in principle, the set of features under investigation is the behavior of each PDF in a certain $x$ region, so in a full analysis, players should be taken as the set of all possible perturbed PDFs, in any $x$ range, and possibly including up and down perturbations. Even if the full physically relevant $x$ range $10^{-4}\lesssim x\lesssim0.7$
was only subdivided into, say, four bins for each PDF, this would lead to $4 N_{\rm flav}$ players, or twice as many if up and down perturbations are included separately, which corresponds to $2^{32}\sim 10^9$ coalitions, thereby making an exact treatment unmanageable. However, we shall show in the next section that even within our simplified setting, the Shapley value performs adequately as a means to probe the impact of data on PDFs.

\section{Analyzing the NNPDF4.0 PDF set}
\label{sec:results}

We now apply the technique discussed in the previous section to the NNPDF4.0 PDF set, with the dual goal of checking the viability of our method and exploring its potentialities. We use the default published NNLO baseline PDF set
\texttt{NNPDF40\_nnlo\_as\_01180}, with 100 PDF replicas, and we adopt the same global dataset used  for the determination of these PDFs, as described in Ref.~\cite{NNPDF:2021njg}.  Individual datasets and experiments included in this global dataset are listed in Table~\ref{tab:dataset_full} of  Appendix~\ref{app:datasets}. The kinematic coverage of this dataset in the $(x,Q^2)$ plane is shown in Fig.~2 of Ref.~\cite{NNPDF:2021njg}.

In the NNPDF4.0  determination (unlike in later NNPDF studies) no theory uncertainty contribution associated with missing perturbative higher orders was included. We consequently adopt as a covariance matrix in the value function Eqs.~(\ref{eq:value-function},\ref{eq:loss_simp_pdf}) the (full) experimental covariance matrix. Note that this differs from the loss used for the training of the neural nets in  Ref.~\cite{NNPDF:2021njg} in  two respects: first, it does not include a contribution introduced to account for nuclear uncertainties in datasets involving nuclear targets, and second, it uses the experimental covariance matrix also for the treatment of multiplicative uncertainties, for which a separate uncertainty treatment is necessary to avoid biased results. 
Consequently, our value function measures the PDF impact on experimental data calibrated to the purely experimental uncertainty.

We will first evaluate the Shapley value for the full dataset, thus assessing the impact of individual PDFs on the description of the global dataset, thereby providing a first check on the viability of the method. We then turn to individual subsets of data, studying how different PDFs are correlated to different processes in specific kinematic regions. Results generally follow textbook~\cite{Ellis:318585,Forte:2013wc} expectations, but we find a few mildly surprising results, and one  specific case in which the Shapley index analysis reveals an unexpected feature of the gluon PDF determination. Finally, we provide two studies testing the potentialities of our methodology. In the first, we show that the unexpected results found in the dataset analysis reveal a potentially dangerous behavior of the gluon PDF uncertainty. Next, we provide a  proof-of-concept application to the design of folds used in the NNPDF hyperparameter optimization procedure.

\subsection{The Shapley value for the  full dataset}
\label{sec:shap_full_data}

We compute the Shapley value for each PDF in turn, using the full global dataset: this provides an estimate of the global impact of PDFs on theory prediction for the NNPDF4.0 dataset.
Before doing this by means of the localized perturbation discussed in Sec.~\ref{sec:setup}, we perform a first check of the method, in which instead of using 
Eq.~(\ref{eq:perturbation-cal}), we simply set
\begin{equation}\label{eq:globzero}
xf_j^{\text{pert}}(x,Q_0^2) = 0.
\end{equation}
This cannot be done for the full NNPDF4.0 dataset, as setting some PDFs to zero leads to the vanishing of some theory predictions, thereby producing ill-defined values for some observables, typically cross-section ratios. We have consequently excluded datasets for which this happens, which are listed in Table~\ref{tab:discarded_datasets_ablation}.

\begin{figure}[t!]
    \centering
    \includegraphics[width=0.99\linewidth]{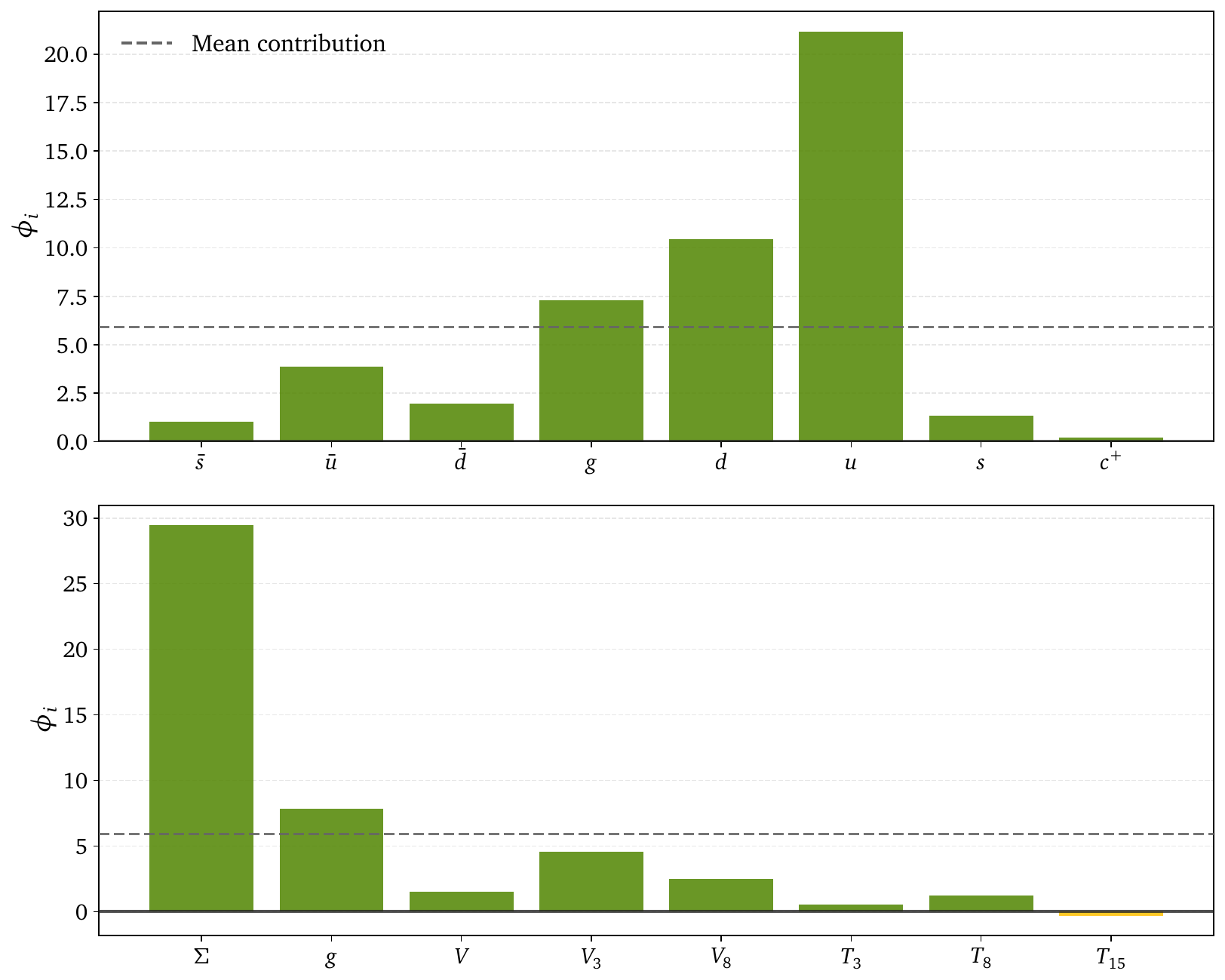}\\
    \caption{Shapley value obtained setting  each PDF to zero in turn according to Eq.~(\ref{eq:globzero}) in the flavor basis (top) and evolution basis (bottom). The dashed horizontal line corresponds to the average over flavors.}
    \label{fig:ablation_sv}
\end{figure}

The Shapley values are shown in Fig.~\ref{fig:ablation_sv} for each PDF in the flavor (top) or evolution (bottom) bases, Eqs.~(\ref{eq:bfldef}),~(\ref{eq:bevdef}), respectively. Comparing the values gives an indication of which PDF has the biggest impact on the theory predictions for the NNPDF4.0 dataset, and conversely, which PDF is most constrained by the NNPDF4.0 data. Note that the loss function used in the computation of the Shapley value is normalized by the experimental uncertainty in the basis of eigenvectors of the experimental covariance matrix, but the perturbation applied to the PDFs is normalized to the PDF  uncertainty. So  what the Shapley value is telling us is the marginal contribution of that PDF to the theory prediction in units of the data uncertainty. 
We find that the flavor-basis Shapley values are ordered according to
\begin{equation}
    \phi_u > \phi_d > \phi_g > \phi_{\bar{u}} > \phi_{\bar{d}} > \phi_s >  \phi_{\bar{s}} > \phi_{c^+},
\end{equation}
in agreement with physical expectations: the up and down quarks, whose valence distributions $u^-$ and $d^-$ carry the proton baryon number (see 
Eq.~\eqref{eq:val_sumrule}), have the largest impact; the gluon, though carrying the largest fraction of the proton momentum, has a lower impact because its contribution to the majority of processes in the NNPDF4.0 dataset starts only at NLO; and heavier quarks have an impact that scales as their reciprocal mass.

\begin{figure}[b!]
    \centering
    \includegraphics[width=0.99\linewidth]{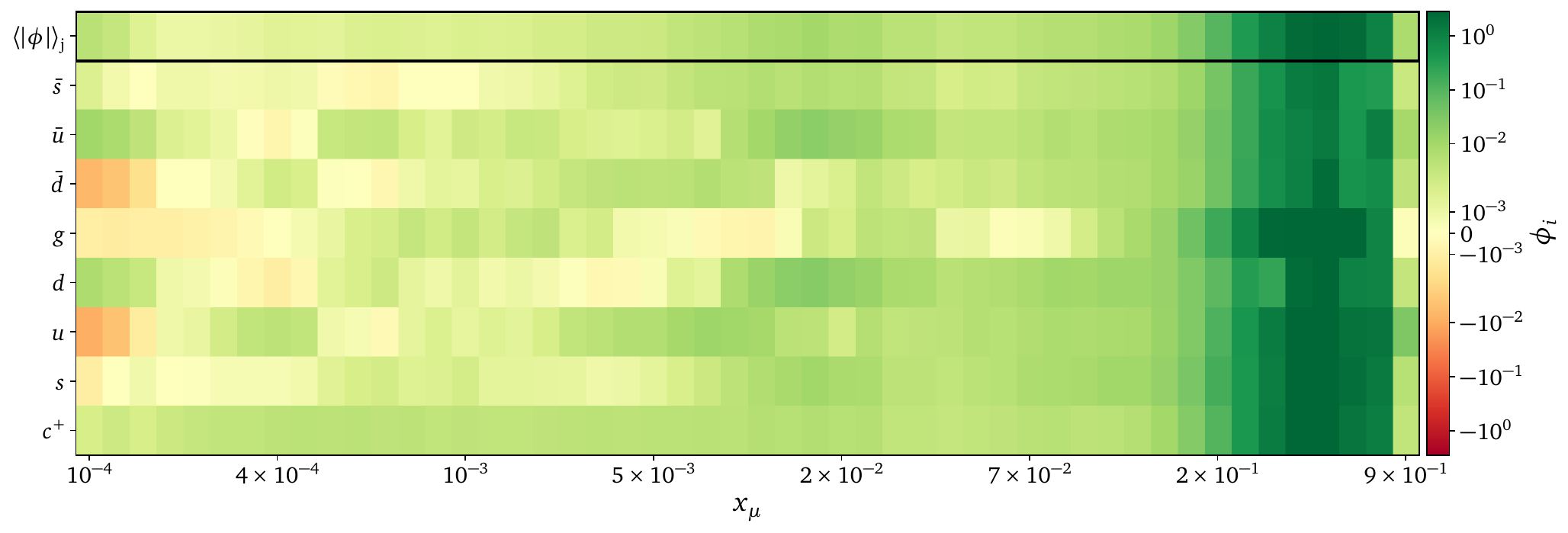}\\
    \includegraphics[width=0.99\linewidth]{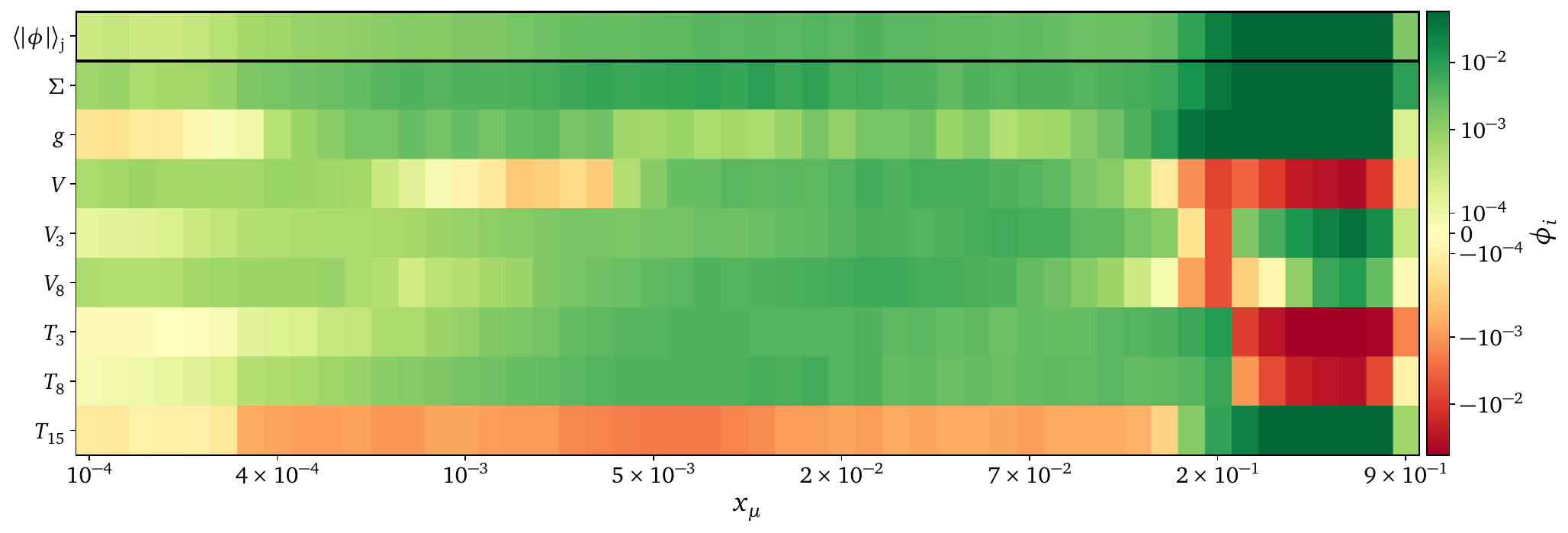}
    \caption{Scan of flavor-basis (top) and evolution-basis (bottom) Shapley values as a function of perturbation center $x_\mu$ for NNPDF4.0, computed for the central PDF set. The value $\phi_j(x_\mu)$ is shown for all PDF flavors $j$ and for 50 values of $x_\mu$, equally spaced on a logarithmic scale in the range $ 10^{-4}\le x_\mu\le 0.9 $. Each column corresponds to one value of $x_\mu$, each row to a parton flavor, and the color encodes the signed contribution magnitude on a symlog scale (green: positive, red: negative). The top row shows the mean absolute contribution across flavors, highlighting the overall sensitivity trend with $x_\mu$.}
    \label{fig:dense-x-range}
\end{figure}

\begin{figure}[t]
    \centering
    \includegraphics[width=.99\linewidth]{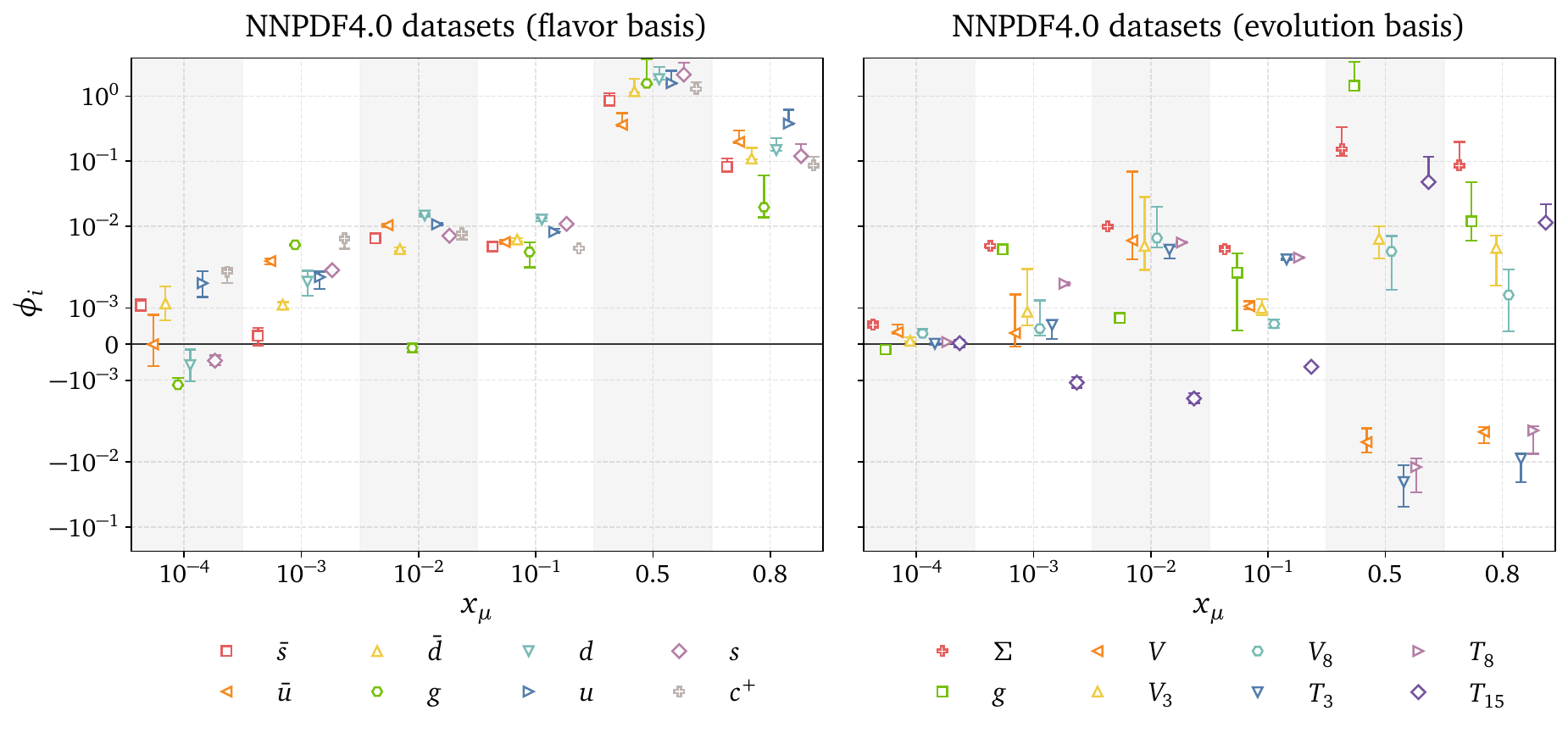}
    \caption{The median Shapley value $\tilde\phi_j(x_\mu)$ over replicas, with an uncertainty band given by the central 68\% interval $[\phi_j^{16},\phi_j^{84}]$ (16th–84th percentiles) of the replica distribution, plotted with a symlog scale on the $y$ axis for six different values of the perturbation center $x_\mu$ for all PDF flavors in the flavor basis (left) and evolution basis (right). The values corresponding to each $x_\mu$ are offset for readability, and are all evaluated at the center of the displayed $x$ bin. The distribution of values over replicas for $x_\mu=0.5$ is displayed in Fig.~\ref{fig:violin}.}
    \label{fig:xrange-nnpdf4.0-pos}
\end{figure}

After this initial test, we turn to the perturbation Eq.~(\ref{eq:perturbation-cal}). We perform a scan across $50$ logarithmically spaced $x_\mu$ values in the interval 
$[10^{-4}, 0.9]$, with $w= \frac{\log_{10}(0.9)-\log_{10}(10^{-4})}{50}$. We first compute  the Shapley index  using the central PDF set: results are shown as heat maps
in Fig.~\ref{fig:dense-x-range} for all values of the perturbation center, both in the flavor and evolution basis. We then compute the values and uncertainties now using Eq.~(\ref{eq:uq}), as a mean and standard deviation over the replica sample. In this case, it turns out that some coalitions lead to pathological values, related again to the vanishing or near-vanishing of some cross sections, which distort the probability distribution, and are discarded using a regularization procedure discussed in Appendix~\ref{app:stabilization}. This procedure will henceforth be applied to all computations of the Shapley value performed using Eq.~(\ref{eq:uq}). Results  
for six different values of the perturbation center (offset to improve readability) are shown in Fig.~\ref{fig:xrange-nnpdf4.0-pos}. 

In order to understand the observed behavior, it is useful to keep in mind that the data are concentrated in the $10^{-3}\lesssim x\lesssim0.5$ region, with the vast majority of data at a scale $Q^2\gtrsim 100$~GeV$^2$ rather higher than the parametrization scale $Q_0$, and that the form of the evolution kernels in Eq.~(\ref{eq:evolution})  is such that the behavior of the initial condition $f_i(x,Q_0^2)$ in some $x\sim x_0$ region mostly affects the behavior of the evolved PDFs $f_i(x,Q^2)$ for $x< x_0$ when the evolved scale is greater than the starting scale $Q>Q_0$. Finally, it should be kept in mind that PDFs are constrained to vanish at the endpoint $x=1$,  and must satisfy the sum rules 
Eqs.~(\ref{eq:mom_sumrule}-\ref{eq:val_sumrule}). 

\begin{figure}[t]
    \centering
    \includegraphics[width=.8\linewidth]{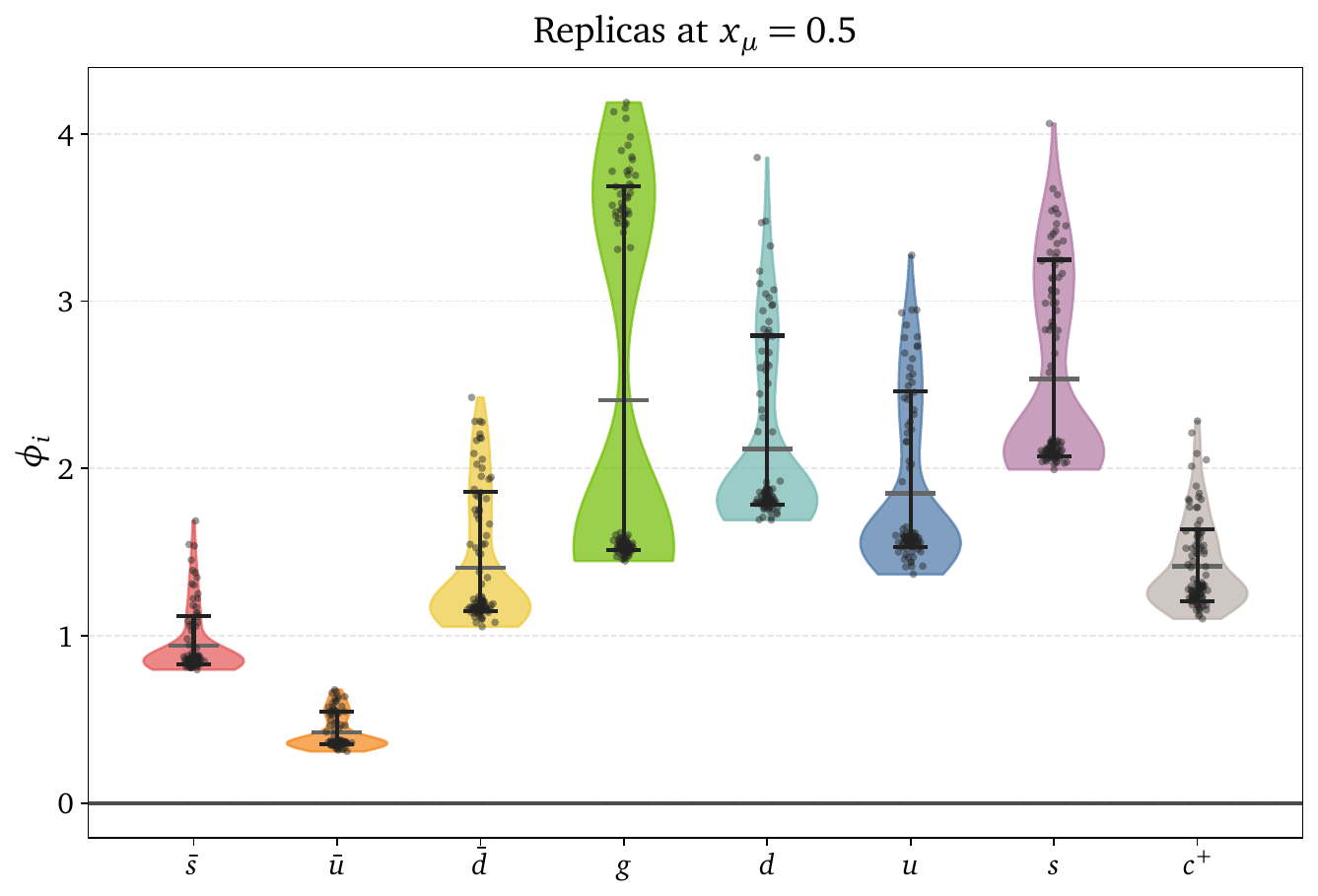}
    \caption{The Shapley values $\phi_j(0.5)$ in the flavor basis (same as shown in Fig.~\ref{fig:xrange-nnpdf4.0-pos}), now shown for all replicas. The width of the colored band corresponds to the density of the replica distribution, while each gray dot corresponds to a single replica, randomly offset for readability. The three solid horizontal lines correspond to the mean, maximum, and minimum over the replica sample. Note that in this plot the  scale on the $y$ axis is  linear.
    }
    \label{fig:violin}
\end{figure}

Based on these considerations, the overall pattern that emerges from Figs.~\ref{fig:dense-x-range}-\ref{fig:xrange-nnpdf4.0-pos} is that Shapley value differs from zero in the data region, and it vanishes for $x\lesssim 10^{-4}$ where there are no data, as it should be. Several more detailed features are less obvious, but can be understood based on the known dependence on PDFs of the partonic cross sections that enter the specific processes in the dataset considered here:
\begin{itemize}
    \item While at very small $x$ the Shapley value fluctuates about zero, one may naively expect a rapid drop at the edge of the data region, rather than the slow decrease observed. Indeed, the perturbation is calibrated to the PDF uncertainty, hence as the data become more sparse at small $x$, the PDF uncertainty increases, the  perturbation grows stronger, and one may consequently expect the Shapley value to remain approximately constant. However, at small $x$ the effect of perturbative evolution according to Eq.~(\ref{eq:evolution}) leads to a rise of the PDF as $x$ decreases and $Q^2$ increases, which is at LO independent of the boundary condition, and thus largely washes out the dependence of evolved PDFs on the boundary condition for $x\lesssim 10^{-2}$ in the region where data are available~\cite{Ball:1994du}. This explains the gradual reduction of sensitivity signaled by the decrease of the Shapley value.
    
    \item The Shapley value is maximum at $x_\mu=0.5$, despite the fact that the majority of the data is at smaller $x$ values. This is driven by the gluon and singlet, and thus it can be understood as a consequence of two effects. First, a large bump in this region then propagates into smaller $x$ regions at higher scales, where data are abundant and precise. Second, because the momentum sum rule is a constraint on the second moment of the PDF, a bump at large $x$ has an enhanced effect on it, and thus leads to sizable deviations in the  whole $x$ range when the normalization is readjusted to maintain the sum rule.
    
    \item The Shapley value remains substantial even for $x_\mu=0.8$, despite the fact that there are almost no data there. This can be understood as a combination of  various effects. First, the bump at $x\sim 0.8$ at the starting scale  will propagate into PDFs at lower $x$ values at the data scale. Furthermore, as mentioned, the perturbation is calibrated so it will become large at large $x$ as the PDF uncertainty grows, and also, the vanishing constraint at $x=1$ forces the PDF uncertainty to shrink to a point as $x\to1$. Finally, the impact of the momentum sum rule is now even larger than at $x_\mu=0.5$.  These observations are interesting in view of searches for heavy resonances that probe PDFs at large $x$, and thus one may be interested in understanding to what extent a signal may or may not be hidden in the PDFs~\cite{Carrazza:2019sec}. This analysis suggests that, despite the lack of direct experimental constraints, the large $x$ PDF uncertainty is reliable enough that a signal of new physics, such as a heavy resonance, would not be easily reabsorbed in it.
    
    \item While the flavor non-singlet valence $V_3$ and $V_8$ combinations are well constrained at large $x\sim 0.3$, the corresponding $q^+$ combinations $T_3$ and $T_8$ are not, and become well constrained only for smaller $x\lesssim 0.2$. This can be understood as a consequence of the fact that flavor non-singlet $q^-_i$ combinations are mostly probed in charged-current interactions, such as $W$ production and neutrino deep-inelastic scattering, covering the whole $x$ range. On the other hand, the flavor non-singlet $q^+$ combinations  $T_3$ and $T_8$  are mostly probed by neutral-current deep-inelastic and fixed-target Drell--Yan data for production on nuclear targets, which are mostly concentrated at smaller $x$.
    
    \item The flavor singlet valence combination $V$ is largely unconstrained at large $x$. This mildly surprising result can be explained by noting that there is in fact no individual data that specifically probes this combination. Specifically, at the LHC, because of the proton--proton initial state, quarks and antiquarks contribute symmetrically to physical observables whose partonic cross section involves a quark--antiquark initial state. As it is clear from a comparison of the flavor-basis and evolution-basis plots in  Fig.~\ref{fig:dense-x-range}, this lack of sensitivity to the total valence combination is only visible when considering evolution-basis PDFs, and not when looking at individual quark flavors, each of which is a combination of singlet, non-singlet and various valence components. 
    This is again relevant for new-physics searches, as it reveals a potential loss of sensitivity to signals that couple to this specific combination and would be masked by the PDF uncertainty.
    
    \item The distributions over replicas shown in Fig.~\ref{fig:violin} display clear deviations from Gaussian behavior, with an asymmetric shape, a  large number of outliers, concentrated in the upper direction, and possibly even a double-peak shape in the case of the gluon. A similar (though somewhat less pronounced) behavior is observed for the other large $x$ values of the perturbation center, $x_\mu=0.1,\,0.8$, but is completely absent for smaller values of $x_\mu$ (not shown here). This can be understood by noting that in the large $x$ region the momentum sum rule constraint has a large impact because of the $x$ weight in Eq.~(\ref{eq:mom_sumrule}), though less so at $x=0.8$ where the PDF is very small. The momentum sum rule is a non-local constraint, hence the effect of the perturbation may affect the prediction for a large number of precisely measured data points, along some direction in   data space. The data replicas to which PDF replicas are fitted are generated as multivariate Gaussian along directions of eigenvectors of the experimental covariance matrix, but, because of the non-linear relation between data and observables, the shifts are non-linearly related to a shift along the direction of the perturbation. Consequently, the effect of the perturbation on individual PDF replicas can be related in a highly non-linear way to their shifts, thereby leading to a non-Gaussian distribution of Shapley values.
\end{itemize}

One particular feature is unexpected: while the Shapley value generally peaks somewhere in the data region and vanishes at very large or very small $x$, where data are absent, the gluon Shapley value instead has a localized dip between $10^{-2}$ and $10^{-1}$, essentially vanishing at $x_\mu\sim0.07$. We discuss this behavior, its likely explanation, and its implications for new-physics searches in Sec.~\ref{sec:gludip} below.

\subsection{Dataset impact}
\label{sec:datasets}

After testing the behavior of the Shapley value over the full dataset, we now turn to studying the impact of individual datasets on PDF determination and, conversely, the impact of individual PDFs on theoretical predictions for specific datasets. To this purpose, we compute the Shapley value for  specific subsets of the full NNPDF4.0 dataset. A complementary exercise would be to compute the Shapley value for data not included in the NNPDF4.0 dataset, but this will be left for future studies. We specifically study deep-inelastic scattering (DIS) data, Drell--Yan, i.e.\ gauge boson production data, and the combination of jet and top pair production data. In each case, we compute the Shapley value for all PDF flavors and for eight values of $x_\mu$ and look at their profile, analogous to Fig.~\ref{fig:xrange-nnpdf4.0-pos}, but with a coarser graining in $x$. When comparing Shapley values across datasets, it must be kept in mind that the value is computed using a perturbation calibrated to the PDF uncertainty, but normalized to the data uncertainty. Consequently, when many different datasets contribute to the determination of an individual PDF, the Shapley values from each dataset will be lower than those found when computing the value for a dataset that entirely or almost entirely determines a PDF. We now discuss each case in turn, by selecting flavor-basis or evolution-basis  plots that highlight the most relevant physics features.

\begin{figure}[t!]
    \includegraphics[width=1\linewidth]{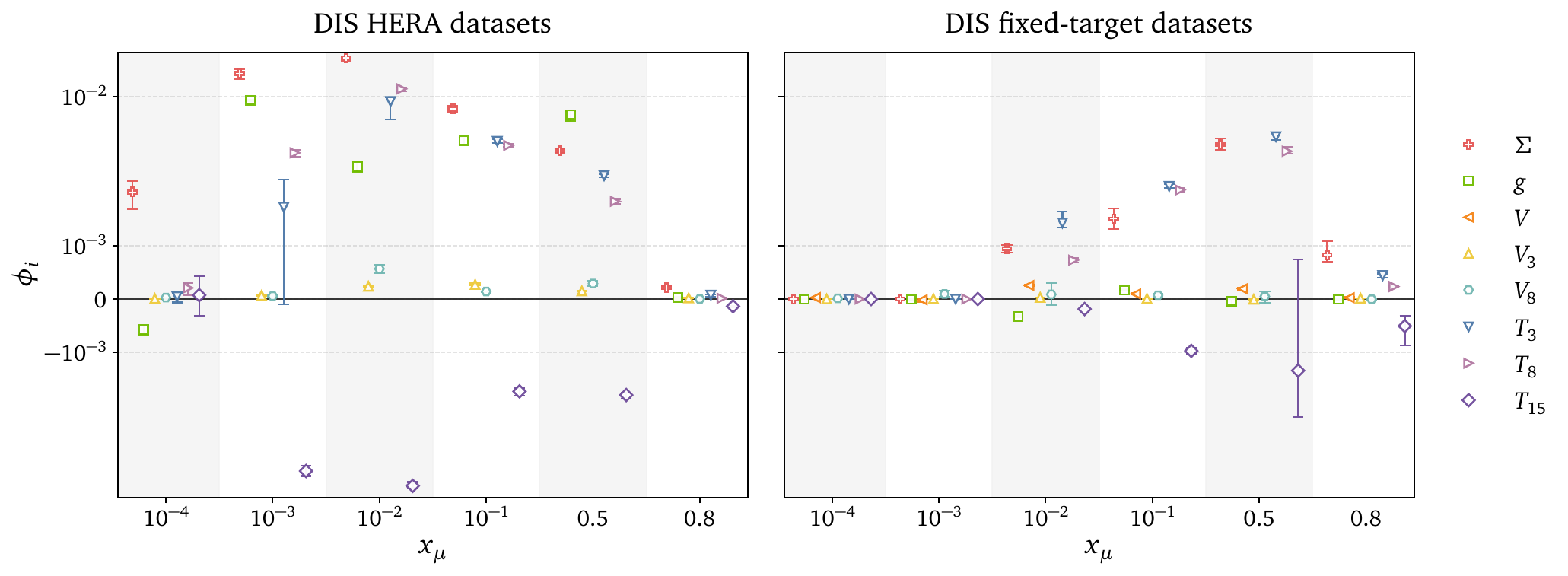}
    \caption{Same as Fig.~\ref{fig:xrange-nnpdf4.0-pos} in the evolution basis, for deep-inelastic scattering data only, from the HERA collider (left) and fixed-target experiments (right). Note that the scale on the $y$ axis differs from that of Fig.~\ref{fig:xrange-nnpdf4.0-pos}.}
    \label{fig:DIS}
\end{figure}

\paragraph{Deep-inelastic scattering}

\begin{figure}[t!]
    \includegraphics[width=1\linewidth]{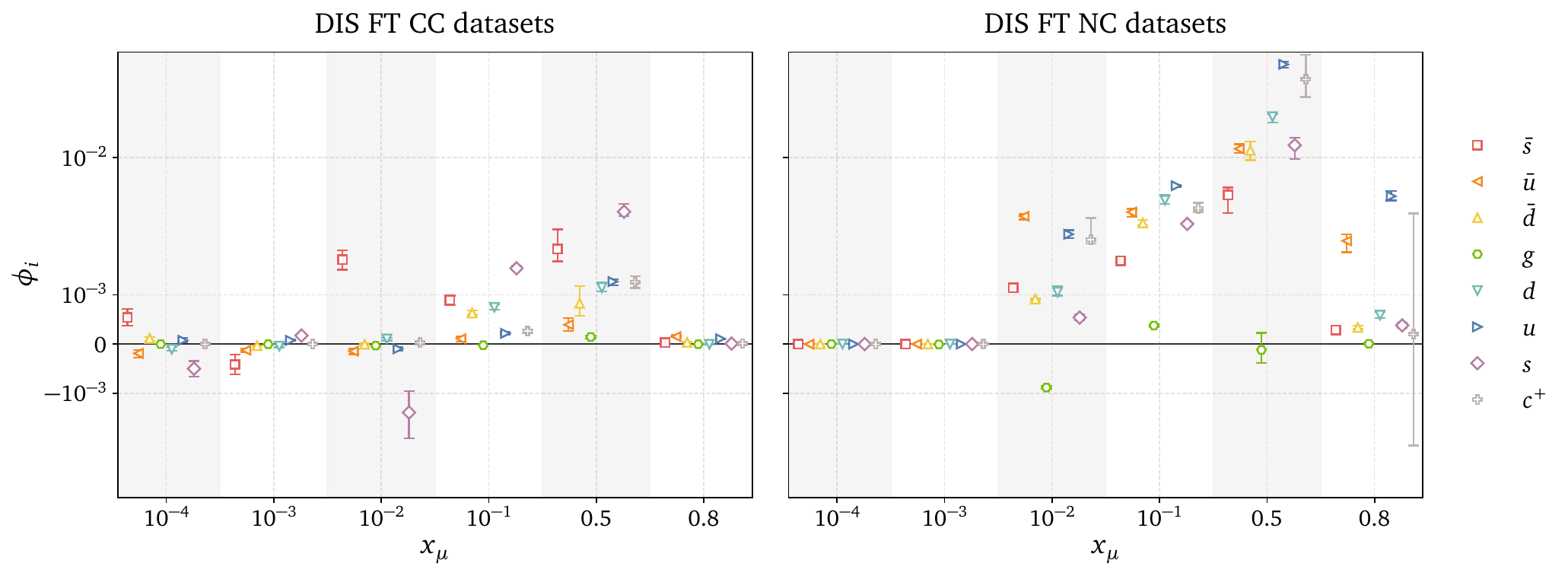}
    \caption{Same as Fig.~\ref{fig:DIS}, but now for fixed-target charged-current (i.e.\ neutrino) DIS data (left)  and neutral-current fixed-target (i.e.\ charged lepton) DIS data (right). Note that the scale on the $y$ axis differs from that of previous plots.}
    \label{fig:CC-vs-NC}
\end{figure}

Deep-inelastic scattering (DIS) data, in principle, allow a full determination of all PDFs if both charged-current (neutrino) and neutral-current (charged lepton) data are available. However, they are only available in a restricted kinematic region. The Shapley value computed  from DIS data for each PDF in the evolution basis is shown in Fig.~\ref{fig:DIS}, where we separate data from the HERA electron--proton collider from data for DIS with electron, muon, and neutrino beams from a variety of fixed-target experiments, and in 
Fig.~\ref{fig:CC-vs-NC}, where we instead compare charged current (neutrino) DIS to neutral current (charged lepton) DIS. The HERA data included in this analysis are listed in Table~\ref{tab:HERA}, and the fixed-target data, separated into charged and neutral current, are listed in Table~\ref{tab:fixed_target}.

The HERA data are at higher energy, hence larger $Q^2$ and smaller $x$ compared to fixed-target data, and their Shapley values are thus peaked at smaller $x$. At small $x\lesssim 0.001$ the singlet and gluon are  larger than any other PDF combination. The values of $Q^2$ reached at HERA are large enough that the contributions from photon and $Z$ exchange  are comparable in the large $Q^2$ bins, which cover the medium-large $x\gtrsim10^{-2}$ region. Consequently the LO DIS cross section depends on three different linear combinations of quark PDFs, so there is also sensitivity to the non-singlet $T_3$ and $T_8$ combinations, which depend on the asymmetry between up, down, and strange $q^+_i$ combinations.

HERA data are available both in a neutral-current and charged-current channel, hence in principle they are also sensitive to valence $q^-_i$ PDF combinations, that depend on the quark--antiquark difference. However, valence PDFs are largely determined by fixed-target neutrino DIS data and Drell--Yan data, and thus the contribution to the Shapley value from HERA data only is very small.  
Fixed-target datasets are only available for larger $x\gtrsim10^{-2}$ and indeed the index correspondingly vanishes at smaller $x$. Because of the lower energy they are quite insensitive to the gluon. Fixed-target DIS data are available both on proton and deuteron targets, thereby accessing the up--down quark difference, and indeed they display very large sensitivity to the triplet combination $T_3$, as well as sensitivity to the independent linear combination $T_8$ that includes strangeness.  As they include neutrino DIS data, that couple asymmetrically to quarks and antiquarks of different flavors, they display some sensitivity to the valence combinations. 

Because valence PDFs are determined by a combination of neutrino and HERA charged-current data and Drell--Yan data, the Shapley value for fixed-target data only remains small. When separating off charged-current from neutral-current data, the smaller Shapley values of charged-current data become visible (see Fig.~\ref{fig:CC-vs-NC}) and the sensitivity of the charged-current data to the valence combinations becomes apparent, along with a sensitivity to the non-singlet combinations $T_3$ and $T_8$, while the sensitivity to the singlet, which is always dominant for neutral-current data, is absent for charged current.

\paragraph{Gauge boson (Drell--Yan) production}

\begin{figure}[t!]
    \centering
    \includegraphics[width=1\linewidth]{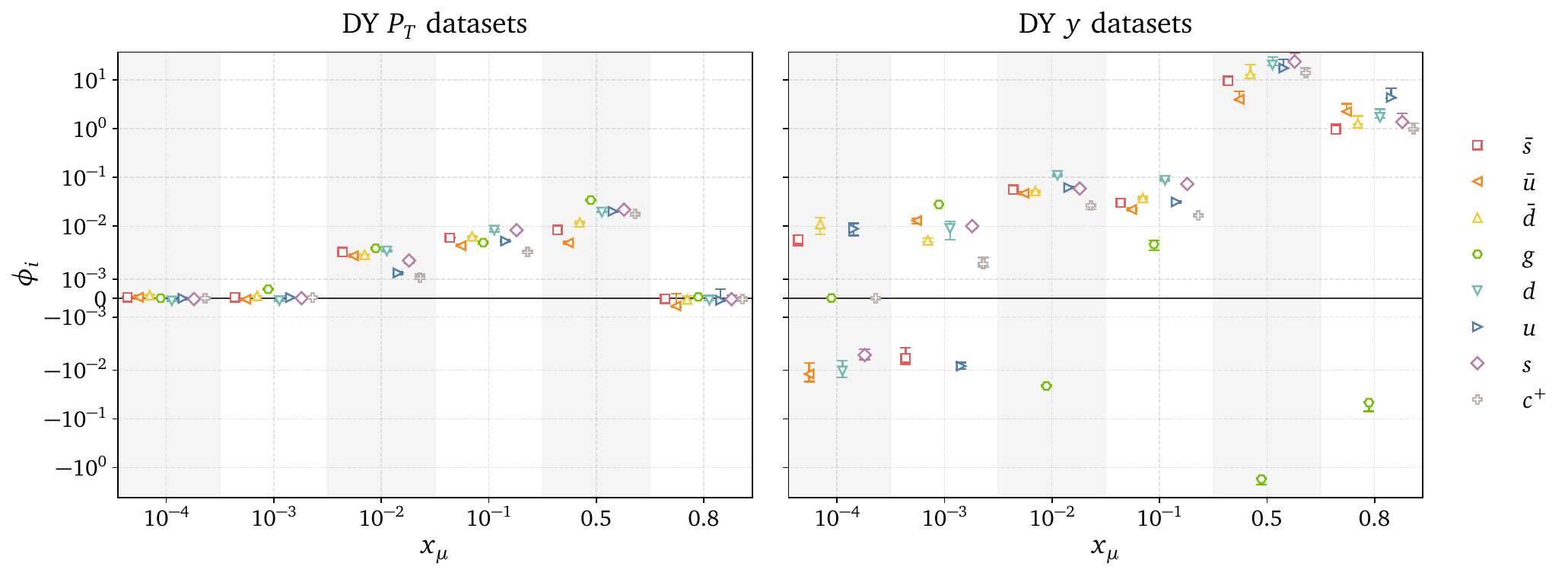}
    \caption{Same as Fig.~\ref{fig:DIS}, but now in the flavor basis, for Drell--Yan transverse momentum (left) and rapidity (right) distributions. Note that the scale on the $y$ axis differs from that of previous plots. }
    \label{fig:DY}
\end{figure}

We now turn to the Drell--Yan process. We consider both transverse momentum and rapidity distributions, with results shown in Fig.~\ref{fig:DY}. The datasets included in either case are listed in Table~\ref{tab:DY_pt} and Table~\ref{tab:DY_rapidity}, respectively. Data on $W$ and $Z$ rapidity distributions from the LHC are extremely precise and abundant over a wide kinematic range, and provide the bulk of the information on flavor separation, especially at very large $x$, as it is clear from the large Shapley values seen in Fig.~\ref{fig:DY} (right) for all quark and antiquark flavors. For smaller $x\lesssim0.1$ values, fixed-target DIS also  plays a significant role, and the Shapley value from Drell--Yan data drops by more than one order of magnitude, becoming comparable to the values seen in Fig.~\ref{fig:CC-vs-NC} for fixed-target DIS.

The Drell--Yan transverse momentum distributions constrain the gluon, that contributes to it at LO, while it only enters the rapidity distribution at NLO. Nevertheless, the gluon Shapley value is comparable for both observables, and in fact even larger for the rapidity distribution for $x \sim 10^{-3}$, for which it is as large as for HERA data, and larger than the Shapley value for quarks. The fact that Drell--Yan rapidity distributions may have a significant impact on the gluon distribution, which is perhaps unexpected, was already observed in Ref.~\cite{fusithesis}.

\paragraph{Top and jets}

\begin{figure}[t!]
    \centering
    \includegraphics[width=1\linewidth]{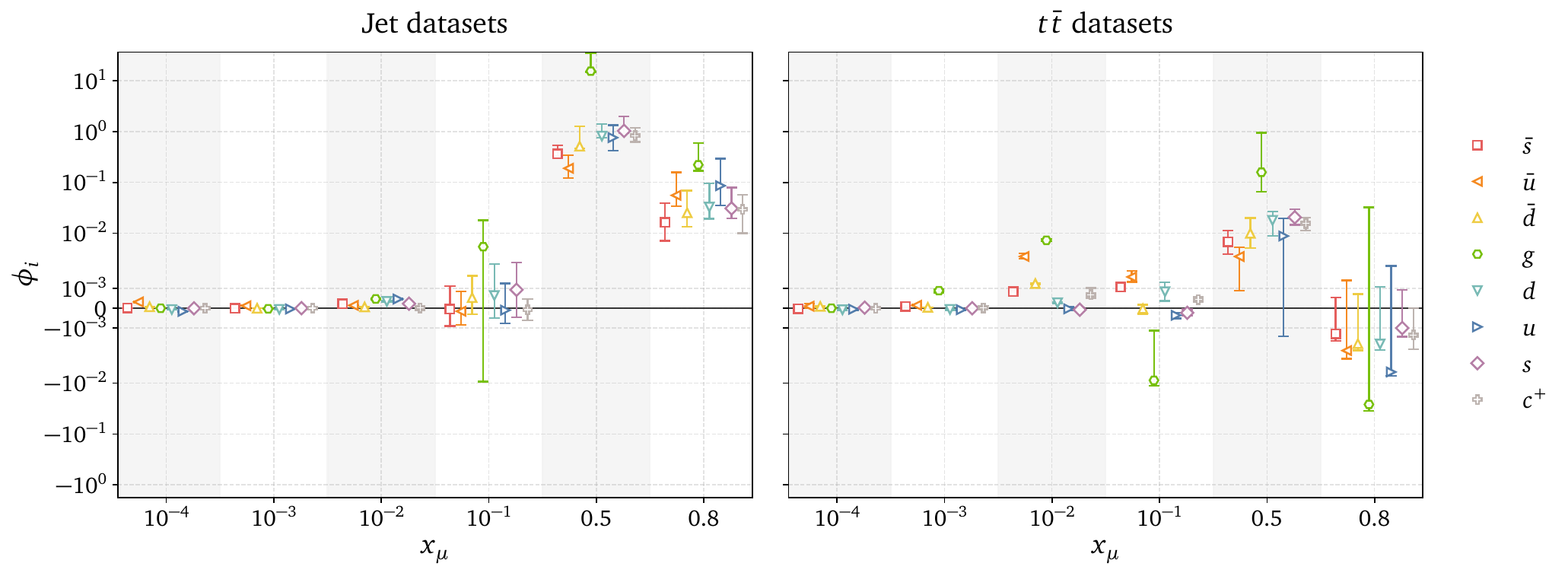}
    \caption{Same as Fig.~\ref{fig:DY}, but now for jet (left) and top pair (right) production. Note that the scale on the $y$ axis differs from that of previous plots. }
    \label{fig:gluon-direct}
\end{figure}

While the gluon PDF at small $x$ is constrained mostly by DIS data, at large $x>0.1$ it is constrained mostly by jet and top pair production data. The Shapley value, computed for the jet and top data respectively listed in Table~\ref{tab:jet} and Table~\ref{tab:tt}, shown in Fig.~\ref{fig:gluon-direct}, confirms this, by displaying a value for the gluon that greatly exceeds that of any other PDF, and moreover a large numerical value, characteristic of the fact that no other data impact the gluon in this region. In the case of top, the LO process comes only from gluon--gluon interactions, and the Shapley index is accordingly negligible for all other PDFs. In the case of jets all PDF flavors contribute on an equal footing, so the Shapley value is significant also for quark PDFs, though in this case many more data, including DIS and Drell--Yan, have a large impact so the value of the index is accordingly smaller.

\subsection{Diagnostics and optimization}
\label{sec:diagopt}

The Shapley value analysis has two main use cases: diagnosing the potentialities and  limitations of a given PDF determination, i.e.\ identifying which PDF flavors 
and $x$ regions are more or less constrained by the data, and optimizing the  PDF determination procedure itself. We present one example of each.

\paragraph{The gluon at intermediate $x$.}
\label{sec:gludip}

In Sec.~\ref{sec:shap_full_data}, we have seen that the gluon Shapley value for the full NNPDF4.0 dataset exhibits a pronounced localized dip in the range $0.05\lesssim x_\mu\lesssim 0.2$, see Fig.~\ref{fig:dense-x-range}. 
This behavior is unexpected: for all other PDF flavors, the Shapley value peaks somewhere in the data region and then decreases gradually towards both small and large $x$. 
The gluon, by contrast, is sizable both below and above this range but drops sharply within it.
To understand this, we recall that the gluon PDF is largely determined by scaling violations, i.e.\ by the $Q^2$-dependence of the data. 
Indeed, for deep-inelastic scattering the gluon only
enters the partonic cross section at NLO, while it already drives the scale dependence at LO through the coupled singlet--gluon evolution
equations, which in Mellin moment space take the form of Eq.~(\ref{eq:evolutionn}), unlike the quark, on which the LO cross section depends.

\begin{figure}[t!]
    \centering
    \includegraphics[width=0.6\linewidth]{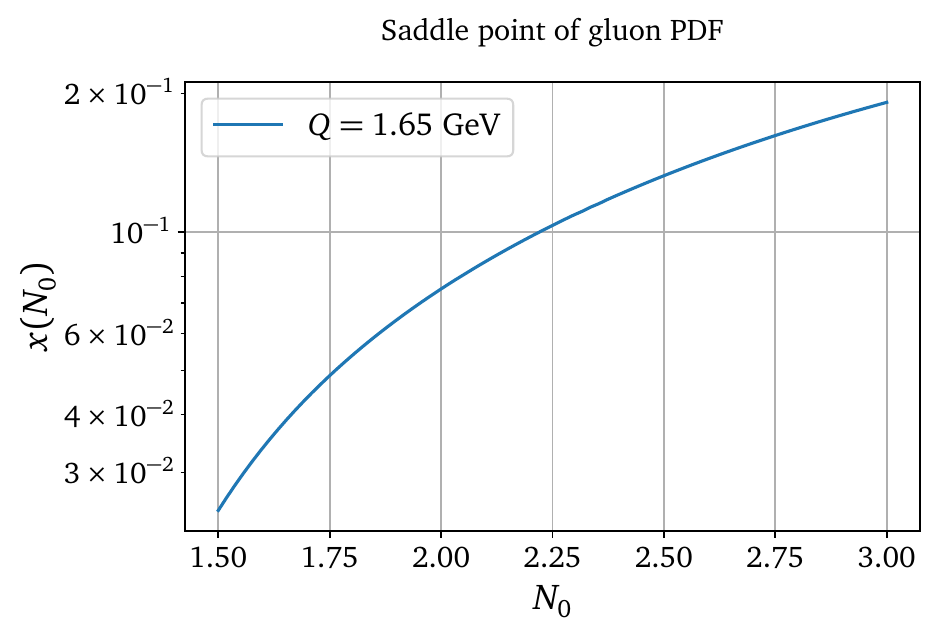}
    \caption{The momentum fraction $x$ corresponding to the  value $N_0$ of the saddle point of the Mellin inversion integral for the gluon distribution, evaluated using NNPDF4.0 PDFs.}
    \label{fig:saddle-point-analysis}
\end{figure}

The key observation is that the momentum sum rule Eq.~(\ref{eq:mom_sumrule}) implies that a specific linear combination of the second Mellin moments ($N=2$) of the quark singlet and the gluon is scale-independent. 
As a consequence, the anomalous dimension matrix $\gamma_{ij}(N)$ in Eq.~(\ref{eq:evolutionn}) has a vanishing eigenvalue at $N=2$: there is an exact flat direction in the space of Mellin-space PDFs, and the corresponding eigenvector -- which is almost aligned with the gluon for $N\lesssim 2$~\cite{Ellis:318585} -- cannot be constrained by scaling violations. 
By continuity, the eigenvalue remains close to zero in a neighborhood of $N=2$, so the gluon is poorly determined by the data for a finite range of Mellin moments around $N=2$.

The question is then how to relate the behavior of the gluon PDF in $N$- and $x$-space.  This can be done easily~\cite{Bonvini:2010tp}, thanks to the fact that the Mellin inversion
\begin{equation}\label{eq:melinv}
     f(x,Q^2) = \frac{1}{2\pi \imag}\int_{c-\imag\infty}^{c+\imag\infty} \dd N\, x^{-N}\,\bar f(N,Q^2).
\end{equation}
that expresses the $x$-space PDF $f(x,Q^2)$ in terms of the $N$-space PDF $\bar f(N,Q^2)$ can be accurately evaluated by saddle point. Writing
\begin{equation}
     f(x,Q^2) = \frac{1}{2\pi \imag}\int_{c-\imag\infty}^{c+\imag\infty}  \dd N \,e^{E(x,N,Q^2)},
     \qquad
     E(x,N,Q^2)\equiv -N\ln x + \ln\bar f(N,Q^2),
\end{equation}
the saddle-point condition $\dd E/\dd N = 0$ gives
\begin{equation}\label{eq:saddle-point}
    \ln x=\frac{\dd\ln\bar f(N)}{\dd N}\Bigg|_{N=N_0},
\end{equation}
which maps each value of $N_0$ to a corresponding $x$ value at which the PDF  $f(x)$ is dominated by $\bar{f}(N_0)$.
In Fig.~\ref{fig:saddle-point-analysis} we show this $x$--$N_0$ correspondence for the NNPDF4.0 gluon evaluated at the reference scale $Q_0$. It is visible that $N_0=2$ maps to $x\approx 7\times10^{-2}$, which coincides with the
center of the dip observed in Fig.~\ref{fig:dense-x-range}. This provides a direct and quantitative explanation: the gluon PDF in $x$-space around $x\sim 0.07$ is
dominated by its Mellin transform near $N=2$, which is the very moment that scaling violations cannot constrain.

One might then expect an inflated gluon uncertainty in this $x$ range. However, this does not happen in practice. For physically motivated functional forms, smoothness of admissible PDF shapes is built in by construction, and it excludes a localized bump. For the neural network parametrization used in NNPDF, a bump is in principle possible, and could be obtained by a targeted training, but it is disfavored by the hyperoptimization of the methodology~\cite{Ball:2022uon}, and only appears in low-probability tails of the PDF distribution. 
In either case, the uncertainty band therefore smoothly interpolates between the neighboring regions without revealing the local loss of sensitivity.
This has a potentially important phenomenological consequence: a genuine localized enhancement of the gluon PDF in this region -- for instance from new physics -- could be absorbed into the fit without producing any visible distortion of the uncertainty band, and would therefore remain hidden from standard PDF diagnostics. 
The Shapley value, by contrast, reveals the insensitivity directly.

\begin{figure}
    \centering
    \includegraphics[width=\linewidth]{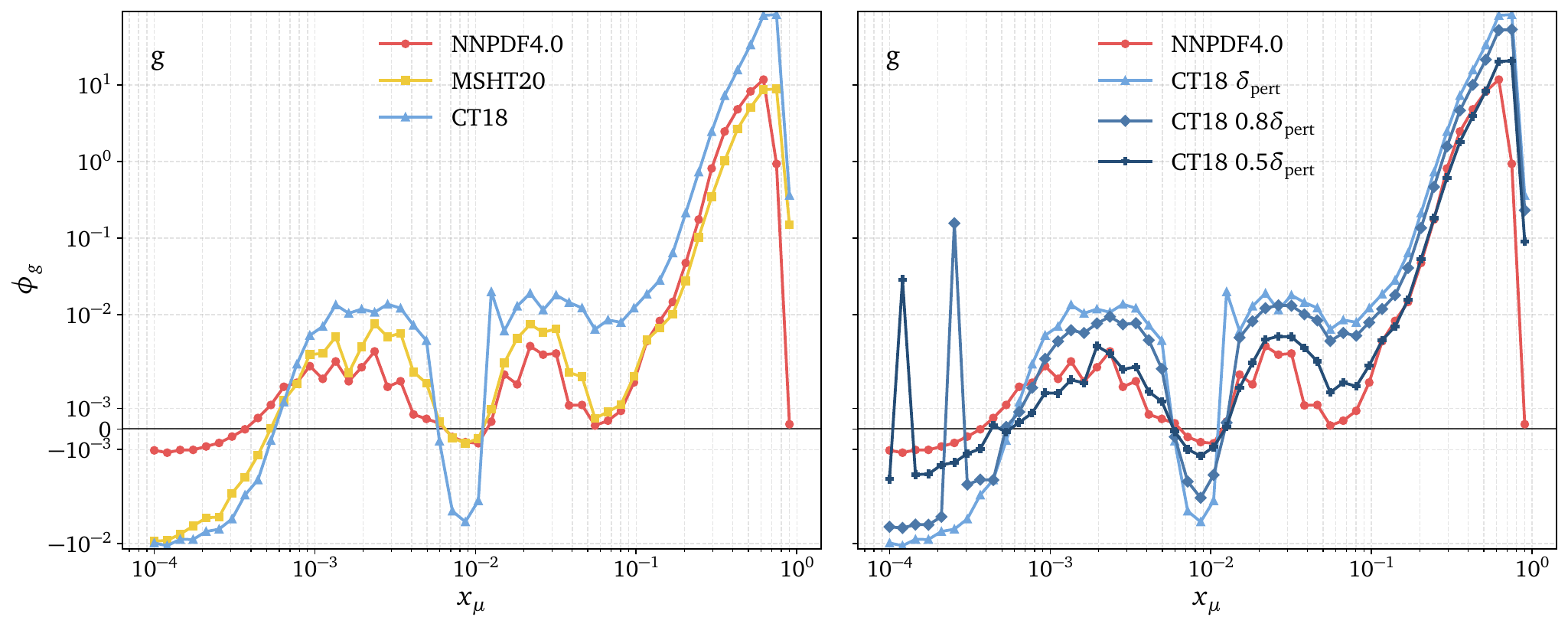}
    \caption{Left: the gluon Shapley value computed from the central NNPDF4.0 PDF (same as shown in Fig.~\ref{fig:dense-x-range}), compared to the results obtained using the MSHT20 and CT18 PDFs. Right: the CT18 gluon Shapley value for various rescalings of the perturbation amplitude, compared to NNPDF4.0.}
    \label{fig:PDFs_set_comparison}
\end{figure}

Because this explanation is based on universal properties of QCD evolution, if correct it should hold for any PDF set. In order to test this, we repeat the analysis using two other global PDF sets, namely CT18~\cite{Hou:2019efy} and MSHT20~\cite{Bailey:2020ooq}, that were compared to NNPDF4.0 in Ref.~\cite{NNPDF:2021njg}: we  apply the same perturbation Eq.~\eqref{eq:perturbation-cal}, calibrated in each case to the uncertainty band of the corresponding set.  We then compute for the gluon of these PDF sets the Shapley values that were shown for all PDFs of the NNPDF4.0 set in Fig.~\ref{fig:dense-x-range}, namely the value obtained using the central PDF and applying the perturbation on  a finely spaced grid.  Also for other PDF sets it turns out that some datasets must be discarded because they lead to pathological behavior; they are also listed in Appendix~\ref{app:stabilization}. 
Results are shown  in Fig.~\ref{fig:PDFs_set_comparison}:  the dip at $x_\mu\approx 7\times10^{-2}$ is present for all three sets. 
While the position of the dip is universal, the overall normalization of the gluon Shapley value varies across sets, as a direct consequence of the perturbation amplitude being set by each set's
own uncertainty band. 
In particular, the CT18 gluon Shapley value lies somewhat above the NNPDF4.0 and MSHT20 results. Rescaling the CT18 perturbation amplitude brings it into close agreement with the other two sets at roughly $0.5\,\delta_{\rm pert}$, demonstrating that the uncertainty estimate for the CT18 PDF set is more conservative than for the MSHT20 and NNPDF4.0 PDF sets, whose uncertainties instead are in broad agreement.

Interestingly, Fig.~\ref{fig:PDFs_set_comparison} also shows a somewhat broader dip for a smaller value $x_\mu\approx 10^{-2}$, which for CT18 is in fact even deeper and extending to negative values. While we see no simple explanation for this dip, it is interesting to observe that it corresponds to the gluon region relevant for $gg\to H$ production
at the LHC. Indeed, the possibility of constructing gluon shapes out of the NNPDF4.0 replicas with a bump in this region, but a very similar $\chi^2$ as the central NNPDF4.0 PDF, was pointed out in Ref.~\cite{Courtoy:2022ocu}. Our analysis shows that a similar behavior would also be observed for MSHT20 and even more for CT18 PDFs by allowing for a PDF parametrization that allows for such a bump.

\paragraph{Fold design}
\label{sec:folds}

\begin{figure}[h!]
\centering
\includegraphics[width=\linewidth]{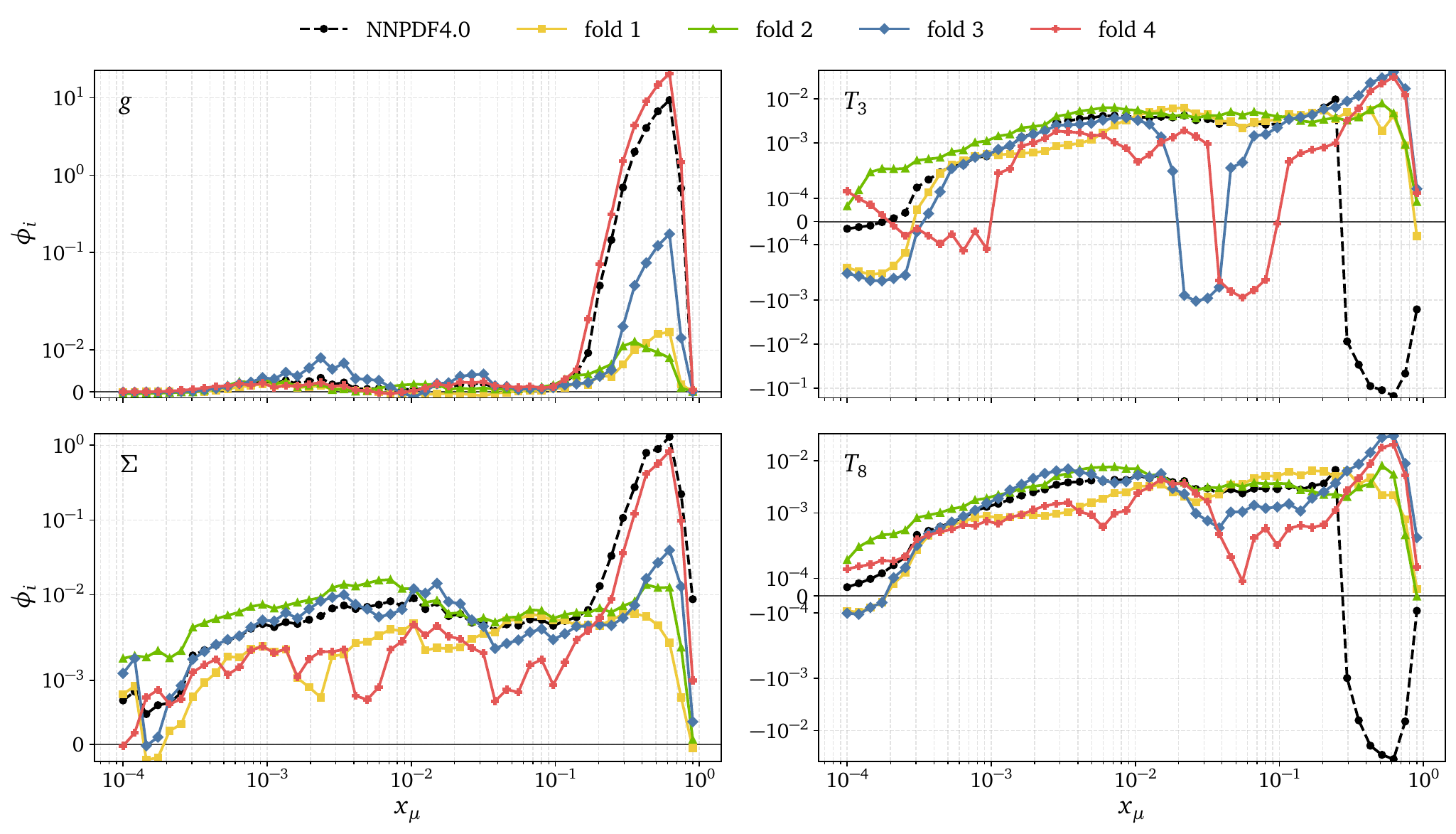} 
\caption{The Shapley value computed using central PDFs (same as in Fig.~\ref{fig:dense-x-range})  now using the full NNPDF4.0 dataset, as well as the datasets corresponding to the four folds used for NNPDF4.0 hyperparameter optimization. Results are shown for the gluon, singlet $\Sigma$, triplet $T_3$ and octet $T_8$ PDFs (see Eqs.~(\ref{eq:singlet},\ref{eq:tdef})).}
\label{fig:folds}
\end{figure}

As discussed in Sec.~\ref{sec:pardet}, the hyperparameters that
control the NNPDF4.0 PDF determination  methodology have been selected
through a hyperoptimization procedure, whereby the data are divided
into four folds. The data contained in each fold should
be representative of the full dataset. The folds actually used in
Ref.~\cite{NNPDF:2021njg} were constructed trying to achieve this
goal, based on physical intuition and knowledge of the properties
of the datasets and physical processes involved. The Shapley value
analysis might provide a better way of verifying the properties of the folds and optimizing their construction.

As a first test of the viability of this idea we have computed the Shapley values as a function of $x_\mu$  for the central NNPDF4.0 PDF for each of the four folds actually used in the NNPDF4.0 hyperoptimization (see Table~8 of Ref.~\cite{NNPDF:2021njg}). 
Results are shown in Fig.~\ref{fig:folds}, where we compare the Shapley values for the gluon, singlet $\Sigma$, triplet $T_3$ and octet $T_8$ PDFs (see Eqs.~(\ref{eq:singlet},\ref{eq:tdef})) computed for each
fold to that of the full dataset. 
Note that, because the values are computed for the central PDF (which is what the  NNPDF4.0 hyperoptimization was based on), these values do not include the PDF uncertainty. 
By and large, the four folds appear to be adequately
designed in that the Shapley profiles are quite similar, showing that the fold design of Ref.~\cite{NNPDF:2021njg} was adequate.
An interesting observation is that at very large $x$, for some PDF combinations, such as the triplet $T_3$ and octet $T_8$, the behavior of the full dataset differs from that of all four folds, with Shapley values becoming negative. This is not contradictory, because the total dataset includes some data that were not included in any of the four folds, as they only became available after the hyperoptimization was performed. The negative Shapley value thus suggests a possible tension between these data and the rest of the dataset.

This first analysis suggests that the Shapley value could be a valuable tool for fold optimization.  
A quantitative way of doing this could be to select the partition of data into folds that minimizes the distance between the Shapley profiles for each PDF in each fold, as measured
for instance by the Kullback--Leibler divergence. 
This could be especially effective given that the more recent NNPDF hyperoptimization methodology~\cite{Cruz-Martinez:2024wiu,Cruz-Martinez:2026aqe}, unlike that of Ref.~\cite{NNPDF:2021njg}, now allows for inclusion of the PDF uncertainty.

\section{Conclusion and outlook}
\label{sec:outlook}

We have introduced a novel, model-agnostic method to interpret global PDF determinations using exact Shapley values. 
Treating the individual PDF flavors as the players of a cooperative game, and the fit likelihood $\chi^2$ as the value function, the Shapley value quantifies the marginal contribution of each flavor, in a given $x$ region, to the agreement between a PDF set and a chosen dataset. 
Because the procedure acts on a previously determined PDF set without reference to how it was obtained, it can be equally applied to ML-based PDF determinations  discussed in Sec.~\ref{sec:black_box}, as well as  to PDFs determined from fixed functional forms. 
The small number of players -- eight flavors, hence $2^7=128$ coalitions per game -- allows the Shapley value to be computed exactly, retaining all correlations between flavors and avoiding the feature-independence approximation that underlies SHAP and is known to distort results in correlated high-energy-physics settings~\cite{Vent:2025ddm}.

Applied to the NNPDF4.0 set, the method reproduces the expected ordering of flavor impact and, scanning over the perturbation center $x_\mu$, it recovers the known map between data and PDFs: sensitivity is concentrated in the data region, washes out at small $x$ where evolution drives the PDF largely independently of its boundary condition, and persists at large $x$ through the combined effect of evolution, the endpoint constraint, and the momentum sum rule. Decomposing the dataset into DIS, Drell--Yan, and jet/top families reproduces the standard lore on which processes constrain which flavors -- HERA fixing the small-$x$ singlet and gluon, fixed-target and charged-current data the large-$x$ non-singlet and valence combinations, Drell--Yan the flavor separation, and jets and top pairs the large-$x$ gluon -- while placing these statements on a quantitative footing.

Beyond confirming expectations, the analysis exposes features that
standard diagnostics do not reveal. The flavor-singlet valence
combination $V$ is found to be essentially unconstrained at large $x$,
reflecting the absence of any single measurement that probes it
directly; this is of direct relevance to searches for new physics
whose quantum numbers might couple to the valence sector. Most strikingly, the gluon Shapley value displays a pronounced dip around $x\sim0.07$, which we trace to the vanishing eigenvalue of the singlet--gluon anomalous-dimension matrix at $N=2$ enforced by momentum conservation, mapped to $x$-space through the saddle point of the Mellin inversion integral. This is a genuine local loss of sensitivity that the smoothness of the parametrization hides inside an otherwise unremarkable uncertainty band, so that a localized enhancement of the gluon in this region could be absorbed by a fit without distorting its errors and would escape standard PDF diagnostics. The Shapley value makes this blind spot explicit. Also, it suggests a potential lack of sensitivity in the region of the gluon PDF which is relevant for Higgs production in gluon fusion at the LHC.

The applications discussed here provide a first proof of concept, and several future applications to actual PDF determination and assessment can be envisaged, keeping in mind that, because the construction takes the PDFs as input and is blind to the way they were determined, it applies unchanged to PDF sets obtained with different methodologies. 
Accordingly, the method can be used for various kinds of PDF diagnostics. 
Specifically, the method enables PDF benchmarking -- such as comparing different PDFs used for precision phenomenology, or constructing combined PDF sets~\cite{PDF4LHCWorkingGroup:2022cjn} -- by providing a direct, methodology-independent measure of where and which PDFs in each set are genuinely constrained by data, as opposed to parametrization or fitting choices. 
A related application is diagnosing the PDF's sensitivity to possible signals of new physics.

On the other hand, it can be used for the assessment of the
potential sensitivity of PDFs to new data, flagging redundant or strongly constraining measurements before they are added to a global determination.
More generally, for fit diagnostics it can be used  to identify
under-constrained regions and localized tensions between
datasets. Finally, it can be used for the optimization of specific
aspects of the PDF determination methodology, such as the $K$-fold
hyperparameter optimization used by NNPDF, where the Shapley index can
be used in order to optimize the partition of data between folds, as
discussed in Sec.~\ref{sec:diagopt}.

More broadly, the construction is not tied to neural networks or to
PDFs at all: because the Shapley value is computed directly on the
mapping from data to predictions, without ever inspecting the internal
structure of the model that produced that mapping, the same approach
applies to any black-box system for which a likelihood or other
suitable reward can be defined. In this respect, our results show how
the Shapley value can be used not only in order to explain the
behavior of a machine learning model, as it is customary in XAI
applications, but also to relate the input and output of a regression
procedure in a model-agnostic way.

An interesting possible line of development is investigating the applicability of our method to more complex sets of features, while
retaining the full exact computation of the Shapley value without
resorting to the additive approximation.
Indeed, we have considered one $x$ region at a time for all PDFs, with a symmetric perturbation, even though in principle all $x$ regions of all PDFs are correlated with each other. A more extensive treatment would therefore require enlarging the feature space to include several $x$ bins for all PDFs simultaneously, together with the sign of the perturbation as an independent feature.
While this would most likely require some approximation in order to treat the very large number of coalitions, it would be interesting to investigate controlled approximations that nonetheless
respect the strong correlations among PDFs. 
We leave this to future work.

\section*{Acknowledgements}
We thank Felix Hekhorn and especially Juan Cruz Martinez for useful comments on a preliminary version of this paper. 
Part of this work was supported by the European Union’s Horizon Europe research and innovation programme under the Marie Sk\l{}odowska-Curie grant agreement No 101168829, Challenging AI with Challenges from Physics: How to solve fundamental problems in Physics by AI and vice versa (AIPHY). Computing resources were provided by the CINECA Leonardo HPC cluster through the INFN grant PML4HEP.

\clearpage
\appendix

\section{Dataset tables}
\label{app:datasets}

We provide the list of datasets used in the  analyses presented in Sec.~\ref{sec:results}. 

{\small
\begin{longtable}{lll}
\toprule
Experiment & Dataset & Reference \\
\midrule
\endfirsthead
\toprule
Experiment & Dataset & Reference \\
\midrule
\endhead
\bottomrule
\multicolumn{3}{r}{\small\textit{(continues on next page)}} \\
\endfoot
\bottomrule
\caption{The NNPDF4.0 dataset. For each dataset we indicate the experiment, the name with which it will be referred to in following tables and the original reference. More details, such as the kinematic coverage and the number of data points, are provided in Tables~1--5 of Ref.~\cite{NNPDF:2021njg}.}
\label{tab:dataset_full}
\endlastfoot
NMC & \texttt{NMC\_NC\_NOTFIXED\_EM-F2} & \cite{NewMuon:1996uwk} \\
NMC & \texttt{NMC\_NC\_NOTFIXED\_P\_EM-SIGMARED} & \cite{NewMuon:1996fwh} \\
SLAC & \texttt{SLAC\_NC\_NOTFIXED\_P\_EM-F2} & \cite{Whitlow:1991uw} \\
SLAC & \texttt{SLAC\_NC\_NOTFIXED\_D\_EM-F2} & \cite{Whitlow:1991uw} \\
BCDMS & \texttt{BCDMS\_NC\_NOTFIXED\_P\_EM-F2} & \cite{BCDMS:1989qop} \\
BCDMS & \texttt{BCDMS\_NC\_NOTFIXED\_D\_EM-F2} & \cite{BCDMS:1989ggw} \\
CHORUS & \texttt{CHORUS\_CC\_NOTFIXED\_PB\_NU-SIGMARED} & \cite{CHORUS:2005cpn} \\
CHORUS & \texttt{CHORUS\_CC\_NOTFIXED\_PB\_NB-SIGMARED} & \cite{CHORUS:2005cpn} \\
NUTEV & \texttt{NUTEV\_CC\_NOTFIXED\_FE\_NU-SIGMARED} & \cite{Mason:2006qa} \\
NUTEV & \texttt{NUTEV\_CC\_NOTFIXED\_FE\_NB-SIGMARED} & \cite{Mason:2006qa} \\
HERA & \texttt{HERA\_NC\_318GEV\_EM-SIGMARED} & \cite{H1:2015ubc} \\
HERA & \texttt{HERA\_NC\_225GEV\_EP-SIGMARED} & \cite{H1:2015ubc} \\
HERA & \texttt{HERA\_NC\_251GEV\_EP-SIGMARED} & \cite{H1:2015ubc} \\
HERA & \texttt{HERA\_NC\_300GEV\_EP-SIGMARED} & \cite{H1:2015ubc} \\
HERA & \texttt{HERA\_NC\_318GEV\_EP-SIGMARED} & \cite{H1:2015ubc} \\
HERA & \texttt{HERA\_CC\_318GEV\_EM-SIGMARED} & \cite{H1:2015ubc} \\
HERA & \texttt{HERA\_CC\_318GEV\_EP-SIGMARED} & \cite{H1:2015ubc} \\
HERA & \texttt{HERA\_NC\_318GEV\_EAVG\_CHARM-SIGMARED} & \cite{H1:2018flt} \\
HERA & \texttt{HERA\_NC\_318GEV\_EAVG\_BOTTOM-SIGMARED} & \cite{H1:2018flt} \\
DYE866 & \texttt{DYE866\_Z0\_800GEV\_DW\_RATIO\_PDXSECRATIO} & \cite{NuSea:2001idv} \\
DYE866 & \texttt{DYE866\_Z0\_800GEV\_PXSEC} & \cite{NuSea:2003qoe} \\
DYE605 & \texttt{DYE605\_Z0\_38P8GEV\_DW\_PXSEC} & \cite{Moreno:1990sf} \\
DYE906 & \texttt{DYE906\_Z0\_120GEV\_DW\_PDXSECRATIO} & \cite{SeaQuest:2021zxb} \\
CDF & \texttt{CDF\_Z0\_1P96TEV\_ZRAP} & \cite{CDF:2010vek} \\
D0 & \texttt{D0\_Z0\_1P96TEV\_ZRAP} & \cite{D0:2007djv} \\
D0 & \texttt{D0\_WPWM\_1P96TEV\_ASY} & \cite{D0:2013xqc} \\
ATLAS & \texttt{ATLAS\_WPWM\_7TEV\_36PB\_ETA} & \cite{ATLAS:2011qdp} \\
ATLAS & \texttt{ATLAS\_Z0\_7TEV\_36PB\_ETA} & \cite{ATLAS:2011qdp} \\
ATLAS & \texttt{ATLAS\_Z0\_7TEV\_49FB\_HIMASS} & \cite{ATLAS:2013xny} \\
ATLAS & \texttt{ATLAS\_Z0\_7TEV\_LOMASS\_M} & \cite{ATLAS:2014ape} \\
ATLAS & \texttt{ATLAS\_WPWM\_7TEV\_46FB\_CC-ETA} & \cite{ATLAS:2016nqi} \\
ATLAS & \texttt{ATLAS\_Z0\_7TEV\_46FB\_CC-Y} & \cite{ATLAS:2016nqi} \\
ATLAS & \texttt{ATLAS\_Z0\_7TEV\_46FB\_CF-Y} & \cite{ATLAS:2016nqi} \\
ATLAS & \texttt{ATLAS\_Z0\_8TEV\_HIMASS\_M-Y} & \cite{ATLAS:2016gic} \\
ATLAS & \texttt{ATLAS\_Z0\_8TEV\_LOWMASS\_M-Y} & \cite{ATLAS:2017rue}\\
ATLAS & \texttt{ATLAS\_Z0\_13TEV\_TOT} & \cite{ATLAS:2016fij} \\
ATLAS & \texttt{ATLAS\_WPWM\_13TEV\_TOT} & \cite{ATLAS:2016fij} \\
ATLAS & \texttt{ATLAS\_WJ\_8TEV\_WP-PT} & \cite{ATLAS:2017irc} \\
ATLAS & \texttt{ATLAS\_WJ\_8TEV\_WM-PT} & \cite{ATLAS:2017irc} \\
ATLAS & \texttt{ATLAS\_Z0J\_8TEV\_PT-M} & \cite{ATLAS:2015iiu} \\
ATLAS & \texttt{ATLAS\_Z0J\_8TEV\_PT-Y} & \cite{ATLAS:2015iiu} \\
ATLAS & \texttt{ATLAS\_TTBAR\_7TEV\_TOT\_X-SEC} & \cite{ATLAS:2014nxi} \\
ATLAS & \texttt{ATLAS\_TTBAR\_8TEV\_TOT\_X-SEC} & \cite{ATLAS:2014nxi} \\
ATLAS & \texttt{ATLAS\_TTBAR\_13TEV\_TOT\_X-SEC} & \cite{ATLAS:2020aln} \\
ATLAS & \texttt{ATLAS\_TTBAR\_8TEV\_LJ\_DIF\_YT-NORM} & \cite{ATLAS:2015lsn} \\
ATLAS & \texttt{ATLAS\_TTBAR\_8TEV\_LJ\_DIF\_YTTBAR-NORM} & \cite{ATLAS:2015lsn} \\
ATLAS & \texttt{ATLAS\_TTBAR\_8TEV\_2L\_DIF\_YTTBAR-NORM} & \cite{ATLAS:2016pal} \\
ATLAS & \texttt{ATLAS\_1JET\_8TEV\_R06\_PTY} & \cite{ATLAS:2017kux} \\
ATLAS & \texttt{ATLAS\_2JET\_7TEV\_R06\_M12Y} & \cite{ATLAS:2013jmu} \\
ATLAS & \texttt{ATLAS\_PH\_13TEV\_XSEC} & \cite{ATLAS:2017nah} \\
ATLAS & \texttt{ATLAS\_SINGLETOP\_7TEV\_TCHANNEL-XSEC} & \cite{ATLAS:2014sxe} \\
ATLAS & \texttt{ATLAS\_SINGLETOP\_13TEV\_TCHANNEL-XSEC} & \cite{ATLAS:2016qhd} \\
ATLAS & \texttt{ATLAS\_SINGLETOP\_7TEV\_T-Y-NORM} & \cite{ATLAS:2014sxe} \\
ATLAS & \texttt{ATLAS\_SINGLETOP\_7TEV\_TBAR-Y-NORM} & \cite{ATLAS:2014sxe} \\
ATLAS & \texttt{ATLAS\_SINGLETOP\_8TEV\_T-RAP-NORM} & \cite{ATLAS:2017rso} \\
ATLAS & \texttt{ATLAS\_SINGLETOP\_8TEV\_TBAR-RAP-NORM} & \cite{ATLAS:2017rso} \\
CMS & \texttt{CMS\_WPWM\_7TEV\_ELECTRON\_ASY} & \cite{CMS:2012ivw} \\
CMS & \texttt{CMS\_WPWM\_7TEV\_MUON\_ASY} & \cite{CMS:2013pzl} \\
CMS & \texttt{CMS\_Z0\_7TEV\_DIMUON\_2D} & \cite{CMS:2013zfg} \\
CMS & \texttt{CMS\_WPWM\_8TEV\_MUON\_Y} & \cite{CMS:2016qqr} \\
CMS & \texttt{CMS\_Z0J\_8TEV\_PT-Y} & \cite{CMS:2015hyl} \\
CMS & \texttt{CMS\_2JET\_7TEV\_M12-Y} & \cite{CMS:2012ftr} \\
CMS & \texttt{CMS\_1JET\_8TEV\_PTY} & \cite{CMS:2016lna} \\
CMS & \texttt{CMS\_TTBAR\_7TEV\_TOT\_X-SEC} & \cite{CMS:2016yys} \\
CMS & \texttt{CMS\_TTBAR\_8TEV\_TOT\_X-SEC} & \cite{CMS:2016yys} \\
CMS & \texttt{CMS\_TTBAR\_13TEV\_TOT\_X-SEC} & \cite{CMS:2015yky} \\
CMS & \texttt{CMS\_TTBAR\_8TEV\_LJ\_DIF\_YTTBAR-NORM} & \cite{CMS:2015rld} \\
CMS & \texttt{CMS\_TTBAR\_5TEV\_TOT\_X-SEC} & \cite{CMS:2017zpm} \\
CMS & \texttt{CMS\_TTBAR\_8TEV\_2L\_DIF\_MTTBAR-YT-NORM} & \cite{CMS:2017iqf} \\
CMS & \texttt{CMS\_TTBAR\_13TEV\_2L\_DIF\_YT} & \cite{CMS:2018adi} \\
CMS & \texttt{CMS\_TTBAR\_13TEV\_LJ\_2016\_DIF\_YT} & \cite{CMS:2018adi}\\
CMS & \texttt{CMS\_SINGLETOP\_7TEV\_TCHANNEL-XSEC} & \cite{CMS:2012xhh} \\
CMS & \texttt{CMS\_SINGLETOP\_8TEV\_TCHANNEL-XSEC} & \cite{CMS:2014mgj} \\
CMS & \texttt{CMS\_SINGLETOP\_13TEV\_TCHANNEL-XSEC} & \cite{CMS:2016lel} \\
LHCB & \texttt{LHCB\_Z0\_7TEV\_DIELECTRON\_Y} & \cite{LHCb:2012gii} \\
LHCB & \texttt{LHCB\_Z0\_8TEV\_DIELECTRON\_Y} & \cite{LHCb:2015kwa} \\
LHCB & \texttt{LHCB\_WPWM\_7TEV\_MUON\_Y} & \cite{LHCb:2015okr} \\
LHCB & \texttt{LHCB\_Z0\_7TEV\_MUON\_Y} & \cite{LHCb:2015okr} \\
LHCB & \texttt{LHCB\_WPWM\_8TEV\_MUON\_Y} & \cite{LHCb:2015mad} \\
LHCB & \texttt{LHCB\_Z0\_8TEV\_MUON\_Y} & \cite{LHCb:2015mad} \\
LHCB & \texttt{LHCB\_Z0\_13TEV\_DIMUON-Y} & \cite{LHCb:2016fbk} \\
LHCB & \texttt{LHCB\_Z0\_13TEV\_DIELECTRON-Y} & \cite{LHCb:2016fbk} \\
\end{longtable}}

\begin{table}[htbp]
\centering
\small
\begin{tabular}{l}
\toprule
 Dataset \\
\midrule
 \texttt{NMC\_NC\_NOTFIXED\_EM-F2}   \\
 \texttt{DYE866\_Z0\_800GEV\_DW\_RATIO\_PDXSECRATIO}   \\
 \texttt{D0\_Z0\_1P96TEV\_ZRAP}   \\
 \texttt{ATLAS\_TTBAR\_8TEV\_LJ\_DIF\_YT-NORM}   \\
 \texttt{ATLAS\_TTBAR\_8TEV\_LJ\_DIF\_YTTBAR-NORM}   \\
 \texttt{ATLAS\_TTBAR\_8TEV\_2L\_DIF\_YTTBAR-NORM}   \\
 \texttt{CMS\_TTBAR\_8TEV\_LJ\_DIF\_YTTBAR-NORM}   \\
 \texttt{CMS\_TTBAR\_8TEV\_2L\_DIF\_MTTBAR-YT-NORM}   \\
 \texttt{ATLAS\_SINGLETOP\_7TEV\_TCHANNEL-XSEC}   \\
 \texttt{ATLAS\_SINGLETOP\_13TEV\_TCHANNEL-XSEC}   \\
 \texttt{ATLAS\_SINGLETOP\_7TEV\_T-Y-NORM}   \\
 \texttt{ATLAS\_SINGLETOP\_7TEV\_TBAR-Y-NORM}   \\
 \texttt{ATLAS\_SINGLETOP\_8TEV\_T-RAP-NORM}   \\
 \texttt{ATLAS\_SINGLETOP\_8TEV\_TBAR-RAP-NORM}   \\
 \texttt{CMS\_SINGLETOP\_8TEV\_TCHANNEL-XSEC}   \\
 \texttt{CMS\_SINGLETOP\_13TEV\_TCHANNEL-XSEC}   \\
 \texttt{D0\_WPWM\_1P96TEV\_ASY}  \\
 \texttt{CMS\_WPWM\_7TEV\_ELECTRON\_ASY}  \\
 \texttt{CMS\_WPWM\_7TEV\_MUON\_ASY}   \\
 \texttt{DYE906\_Z0\_120GEV\_DW\_PDXSECRATIO}  \\
\bottomrule
\end{tabular}
\caption{Datasets excluded from the Shapley value computation of Fig.~\ref{fig:ablation_sv}.}
\label{tab:discarded_datasets_ablation}
\end{table}

\begin{table}[htbp]
\centering
\small
\begin{tabular}{ll}
\toprule
 Dataset & Observable \\
\midrule
\texttt{HERA\_NC\_318GEV\_EM-SIGMARED} & NC DIS \\
 \texttt{HERA\_NC\_225GEV\_EP-SIGMARED} & NC DIS \\
 \texttt{HERA\_NC\_251GEV\_EP-SIGMARED} & NC DIS \\
 \texttt{HERA\_NC\_300GEV\_EP-SIGMARED} & NC DIS \\
 \texttt{HERA\_NC\_318GEV\_EP-SIGMARED} & NC DIS \\
 \texttt{HERA\_CC\_318GEV\_EP-SIGMARED} & CC DIS \\
 \texttt{HERA\_CC\_318GEV\_EM-SIGMARED} & CC DIS \\
 \texttt{HERA\_NC\_318GEV\_EAVG\_CHARM-SIGMARED} & charm DIS \\
 \texttt{HERA\_NC\_318GEV\_EAVG\_BOTTOM-SIGMARED} & bottom DIS \\
\bottomrule
\end{tabular}
\caption{The  HERA DIS data.}
\label{tab:HERA}
\end{table}

\begin{table}[htbp]
\centering
\small
\begin{tabular}{lll}
\toprule
Dataset & Observable \\
\midrule
 \texttt{NMC\_NC\_NOTFIXED\_EM-F2} & NC DIS \\
\texttt{NMC\_NC\_NOTFIXED\_P\_EM-SIGMARED} & NC DIS \\
 \texttt{SLAC\_NC\_NOTFIXED\_P\_EM-F2} & NC DIS \\
 \texttt{SLAC\_NC\_NOTFIXED\_D\_EM-F2} & NC DIS \\
 \texttt{BCDMS\_NC\_NOTFIXED\_P\_EM-F2} & NC DIS \\
 \texttt{BCDMS\_NC\_NOTFIXED\_D\_EM-F2} & NC DIS \\
 \texttt{CHORUS\_CC\_NOTFIXED\_PB\_NU-SIGMARED} & CC DIS \\
 \texttt{CHORUS\_CC\_NOTFIXED\_PB\_NB-SIGMARED} & CC DIS \\
 \texttt{NUTEV\_CC\_NOTFIXED\_FE\_NU-SIGMARED} & CC DIS \\
 \texttt{NUTEV\_CC\_NOTFIXED\_FE\_NB-SIGMARED} & CC DIS \\
\bottomrule
\end{tabular}
\caption{The fixed-target DIS data.}
\label{tab:fixed_target}
\end{table}

\begin{table}[htbp]
\centering
\small
\begin{tabular}{ll}
\toprule
Dataset & Observable \\
\midrule
\texttt{ATLAS\_WJ\_8TEV\_WP-PT} & $W^+$ transverse momentum \\
\texttt{ATLAS\_WJ\_8TEV\_WM-PT} & $W^-$ transverse momentum \\
\texttt{ATLAS\_Z0J\_8TEV\_PT-M} & $Z$ $p_T$ vs invariant mass \\
\texttt{ATLAS\_Z0J\_8TEV\_PT-Y} & $Z$ $p_T$ vs rapidity \\
\texttt{CMS\_Z0J\_8TEV\_PT-Y} & $Z$ $p_T$ vs rapidity \\
\bottomrule
\end{tabular}
\caption{Drell--Yan transverse momentum distribution datasets.}
\label{tab:DY_pt}
\end{table}

\begin{table}[htbp]
\centering
\small
\begin{tabular}{ll}
\toprule
Dataset & Observable \\
\midrule
\texttt{CDF\_Z0\_1P96TEV\_ZRAP} & $Z$ rapidity \\
\texttt{D0\_Z0\_1P96TEV\_ZRAP} & $Z$ rapidity \\
\texttt{ATLAS\_WPWM\_7TEV\_36PB\_ETA} & $W^\pm$ pseudorapidity \\
\texttt{ATLAS\_Z0\_7TEV\_36PB\_ETA} & $Z$ pseudorapidity \\
\texttt{ATLAS\_WPWM\_7TEV\_46FB\_CC-ETA} & $W^\pm$ pseudorapidity (central) \\
\texttt{ATLAS\_Z0\_7TEV\_46FB\_CC-Y} & $Z$ rapidity (central-central) \\
\texttt{ATLAS\_Z0\_7TEV\_46FB\_CF-Y} & $Z$ rapidity (central-forward) \\
\texttt{ATLAS\_Z0\_8TEV\_HIMASS\_M-Y} & $Z$ rapidity (high mass) \\
\texttt{ATLAS\_Z0\_8TEV\_LOWMASS\_M-Y} & $Z$ rapidity (low mass) \\
\texttt{CMS\_WPWM\_8TEV\_MUON\_Y} & $W^\pm$ rapidity \\
\texttt{LHCB\_Z0\_7TEV\_DIELECTRON\_Y} & $Z$ rapidity ($e^+e^-$) \\
\texttt{LHCB\_Z0\_8TEV\_DIELECTRON\_Y} & $Z$ rapidity ($e^+e^-$) \\
\texttt{LHCB\_WPWM\_7TEV\_MUON\_Y} & $W^\pm$ rapidity ($\mu$) \\
\texttt{LHCB\_Z0\_7TEV\_MUON\_Y} & $Z$ rapidity ($\mu^+\mu^-$) \\
\texttt{LHCB\_WPWM\_8TEV\_MUON\_Y} & $W^\pm$ rapidity ($\mu$) \\
\texttt{LHCB\_Z0\_8TEV\_MUON\_Y} & $Z$ rapidity ($\mu^+\mu^-$) \\
\texttt{LHCB\_Z0\_13TEV\_DIMUON-Y} & $Z$ rapidity ($\mu^+\mu^-$) \\
\texttt{LHCB\_Z0\_13TEV\_DIELECTRON-Y} & $Z$ rapidity ($e^+e^-$) \\
\bottomrule
\end{tabular}
\caption{Drell--Yan rapidity ($y$) and pseudorapidity ($\eta$) distribution datasets.}
\label{tab:DY_rapidity}
\end{table}

\begin{table}[htbp]
\centering
\small
\begin{tabular}{ll}
\toprule
Dataset & Observable \\
\midrule
\texttt{ATLAS\_1JET\_8TEV\_R06\_PTY} & inclusive jets \\
\texttt{ATLAS\_2JET\_7TEV\_R06\_M12Y} & dijets \\
\texttt{CMS\_2JET\_7TEV\_M12-Y} & dijets \\
\texttt{CMS\_1JET\_8TEV\_PTY} & inclusive jets \\
\bottomrule
\end{tabular}
\caption{Single-inclusive jet and dijet datasets.}
\label{tab:jet}
\end{table}

\begin{table}[htbp]
\centering
\small
\begin{tabular}{ll}
\toprule
Dataset & Observable \\
\midrule
\texttt{ATLAS\_TTBAR\_7TEV\_TOT\_X-SEC} & total $t\bar{t}$ \\
\texttt{ATLAS\_TTBAR\_8TEV\_TOT\_X-SEC} & total $t\bar{t}$ \\
\texttt{ATLAS\_TTBAR\_13TEV\_TOT\_X-SEC} & total $t\bar{t}$ \\
\texttt{CMS\_TTBAR\_5TEV\_TOT\_X-SEC} & total $t\bar{t}$ \\
\texttt{CMS\_TTBAR\_7TEV\_TOT\_X-SEC} & total $t\bar{t}$ \\
\texttt{CMS\_TTBAR\_8TEV\_TOT\_X-SEC} & total $t\bar{t}$ \\
\texttt{CMS\_TTBAR\_13TEV\_TOT\_X-SEC} & total $t\bar{t}$ \\
\texttt{ATLAS\_TTBAR\_8TEV\_LJ\_DIF\_YT-NORM} & $d\sigma/dy_t$ \\
\texttt{ATLAS\_TTBAR\_8TEV\_LJ\_DIF\_YTTBAR-NORM} & $d\sigma/dy_{t\bar{t}}$ \\
\texttt{ATLAS\_TTBAR\_8TEV\_2L\_DIF\_YTTBAR-NORM} & $d\sigma/dy_{t\bar{t}}$ \\
\texttt{CMS\_TTBAR\_8TEV\_LJ\_DIF\_YTTBAR-NORM} & $d\sigma/dy_{t\bar{t}}$ \\
\texttt{CMS\_TTBAR\_8TEV\_2L\_DIF\_MTTBAR-YT-NORM} & $d\sigma/dm_{t\bar{t}}$ \\
\texttt{CMS\_TTBAR\_13TEV\_2L\_DIF\_YT} & $d\sigma/dy_t$ \\
\texttt{CMS\_TTBAR\_13TEV\_LJ\_2016\_DIF\_YT} & $d\sigma/dy_t$ \\
\bottomrule
\end{tabular}
\caption{Top pair production datasets.}
\label{tab:tt}
\end{table}

\clearpage
\section{Dataset regularization.}
\label{app:stabilization}

For some Shapley value computations, the calibrated perturbation of Eq.~(\ref{eq:perturbation-cal}) may lead to unphysical vanishing or near-vanishing of some cross sections, which in turn leads to diverging values for observables that correspond to cross-section ratios. This behavior biases the entire Shapley game, leading to unreliable values of the Shapley central value and uncertainty. To prevent this, we iteratively remove any dataset where $v_{\mathcal{D}_i}(S)>10^5$, where $\mathcal{D}_i\subset\mathcal{D}$ is the $i$-th subset of $\mathcal{D}$, and re-evaluate until no dataset is flagged by this condition. Across all analyses performed for this work, this only happens for the two fixed-target Drell--Yan deuteron/proton ratios of Refs.~\cite{NuSea:2001idv,SeaQuest:2021zxb},
for particular values of the perturbation center $x_\mu$. The discarded datasets are listed  in Table~\ref{tab:discarded_stabilization}, along with the PDF sets and values of the perturbation center for which this regularization procedure is necessary.

\begin{table}[htbp]
\centering
\small
\begin{tabular}{lllp{0.42\linewidth}}
\toprule
Figure & PDF set & Basis & Discarded dataset and centers $x_\mu$ \\
\midrule
Fig.~\ref{fig:dense-x-range}       & NNPDF4.0 & evolution & \texttt{DYE906}: $0.30,\,0.36,\,0.43,\,0.52,\,0.62,\,0.75$ \\
Fig.~\ref{fig:xrange-nnpdf4.0-pos} & NNPDF4.0 & flavor    & \texttt{DYE906}: $10^{-3},\,10^{-2},\,0.5$ \\
Fig.~\ref{fig:xrange-nnpdf4.0-pos} & NNPDF4.0 & evolution & \texttt{DYE906}: $0.1,\,0.5,\,0.8$ \\
\midrule
Fig.~\ref{fig:PDFs_set_comparison} & CT18  & flavor & \texttt{DYE906}: $28$ $x_\mu$, $10^{-4}\!-\!2.2\times10^{-2}$; \texttt{DYE866}: $15$ $x_\mu$, $10^{-4}\!-\!3.4\times10^{-3}$ \\
Fig.~\ref{fig:PDFs_set_comparison} & CT18 ($0.8\delta_{\text{pert}}$) & flavor & \texttt{DYE906}: $26$ $x_\mu$, $10^{-4}\!-\!1.3\times10^{-2}$; \texttt{DYE866}: $9$ $x_\mu$, $1.5\times10^{-4}\!-\!1.6\times10^{-3}$ \\
Fig.~\ref{fig:PDFs_set_comparison} & CT18 ($0.5\delta_{\text{pert}}$) & flavor & \texttt{DYE906}: $17$ $x_\mu$, $10^{-4}\!-\!4.1\times10^{-3}$; \texttt{DYE866}: $1.5\times10^{-4}$ \\
Fig.~\ref{fig:PDFs_set_comparison} & MSHT20 & flavor & \texttt{DYE906}: $15$ $x_\mu$, $10^{-4}\!-\!2.4\times10^{-3}$; \texttt{DYE866}: $1.8,\,2.1,\,3.7\times10^{-4}$ \\
\bottomrule
\end{tabular}
\caption{Datasets discarded by the iterative stabilization procedure. In each case we list the  figure displaying the results of the corresponding Shapley value computation, the PDF set, the PDF basis, and dataset and  perturbation center. Here $\texttt{DYE906}$ denotes the $\texttt{DYE906\_Z0\_120GEV\_DW\_PDXSECRATIO}$ set and  $\texttt{DYE866}$ the $\texttt{DYE866\_Z0\_800GEV\_DW\_RATIO\_PDXSECRATIO}$ set of Table~\ref{tab:dataset_full}. For CT18 and MSHT20 we provide the number of discarded $x_\mu$ values and the range.}
\label{tab:discarded_stabilization}
\end{table}

\clearpage

\bibliographystyle{UTPstyle}
\bibliography{ML_ref,PDF_ref,data_ref}
\end{document}